%% file: main.tex
\documentclass[conference,compsoc]{IEEEtran}
\IEEEoverridecommandlockouts

\usepackage{subcaption}
\usepackage{tcolorbox} 
\usepackage{multirow}
\usepackage{booktabs}
\usepackage{graphicx}
\usepackage{soul}
\usepackage{textcomp}
\usepackage{lipsum}
\usepackage{algorithm}
\usepackage{xspace}
\usepackage[noend]{algpseudocode}
\usepackage{colortbl}

\usepackage{verbatim}
\usepackage[export]{adjustbox}
\usepackage{mdframed}
\usepackage{tikz}
\usepackage{listings}
\usepackage{array}
\usepackage{scalerel,stackengine}
\usepackage{hyphenat}
\usepackage{pifont}
\usepackage{amsmath}
\usepackage{siunitx}
\usepackage{makecell}
\usepackage{circledsteps}
\usepackage{balance}
\usepackage{tabularx}
\usepackage{color}
\usepackage{paralist}
\usepackage{enumitem}
\usepackage{caption}

\usepackage{cite}
\usepackage[hidelinks]{hyperref}
\usepackage{xurl}
\usepackage[T1]{fontenc}

\hypersetup{
    pdftitle={A Queryable Graph-Based Security Analysis Framework for O-RAN},
    pdfkeywords={O-RAN, 5G security, threat modeling, graph database, security analysis}
}

\definecolor{mGreen}{rgb}{0,0.6,0}
\definecolor{mGray}{rgb}{0.5,0.5,0.5}
\definecolor{mPurple}{rgb}{0.58,0,0.82}
\definecolor{backgroundColour}{rgb}{1,1,1}
\definecolor{lightgreen}{rgb}{0.8,1.0,0.8}
\definecolor{darkgreen}{rgb}{0.0, 0.5, 0.0}

\newcommand{\etal}[2]{#1 {et al.} #2}

\newcommand{\codot}
{\protect\tikz[baseline=-0.5ex]\protect\fill[orange!50!yellow] (0,0) circle (1.ex);}

\newcommand{\idot}
{\protect\tikz[baseline=-0.5ex]\protect\fill[violet!50] (0,0) circle (1.ex);}

\newcommand{\ordot}
{\protect\tikz[baseline=-0.5ex]\protect\fill[orange!90] (0,0) circle (1.ex);}

\newcommand{\lbdot}
{\protect\tikz[baseline=-0.5ex]\protect\fill[cyan!50] (0,0) circle (1.ex);}

\newcommand{\dbdot}
{\protect\tikz[baseline=-0.5ex]\protect\fill[blue!60] (0,0) circle (1.ex);}

\lstdefinestyle{CStyle}{
    backgroundcolor=\color{backgroundColour},
    commentstyle=\color{mGreen},
    keywordstyle=\color{mGreen},
    stringstyle=\color{mPurple},
    numberstyle=\tiny\color{mGray},
    basicstyle=\scriptsize\ttfamily,
    breakatwhitespace=false,
    breaklines=true,
    captionpos=b,
    keepspaces=true,
    numbers=none,
    numbersep=5pt,
    showspaces=false,
    showstringspaces=false,
    showtabs=false,
    tabsize=2,
    upquote=true,
    language=SQL,
    aboveskip=1pt,
    belowskip=1pt,
    abovecaptionskip=1pt,
    belowcaptionskip=-6pt,
}

\lstdefinestyle{Text}{
    backgroundcolor=\color{backgroundColour},
    commentstyle=\color{mGreen},
    keywordstyle=\color{mGreen},
    stringstyle=\color{black},
    numberstyle=\tiny\color{mGray},
    basicstyle=\scriptsize\ttfamily,
    breakatwhitespace=false,
    breaklines=true,
    captionpos=b,
    keepspaces=true,
    numbers=none,
    numbersep=5pt,
    showspaces=false,
    showstringspaces=false,
    showtabs=false,
    tabsize=2,
    upquote=true,
    language=SQL,
    moredelim=**[is][\color{diffrem}]{\\-}{\\-},
    moredelim=**[is][\color{darkgreen}]{\\+}{\\+},
    moredelim=**[is][\bfseries]{\\b}{\\b},
    moredelim=**[is][\mdseries]{\\ub}{\\ub},
    moredelim=**[is][\color{black}]{\\uc}{\\uc},
    aboveskip=1pt,
    belowskip=1pt,
    abovecaptionskip=1pt,
    belowcaptionskip=-6pt,
}

\let\OLDthebibliography\thebibliography
\renewcommand\thebibliography[1]{
  \OLDthebibliography{#1}
  \setlength{\parskip}{0pt}
  \setlength{\itemsep}{0pt plus 0.3ex} %
}

\begin{document}

\title{A Queryable Graph-Based Security Analysis Framework for O-RAN}

\author{
\IEEEauthorblockN{
Corban Villa\IEEEauthorrefmark{1},
Michele Guerra\IEEEauthorrefmark{2},
Syed Khandker\IEEEauthorrefmark{2},
Evangelos Bitsikas\IEEEauthorrefmark{3},
Aanjhan Ranganathan\IEEEauthorrefmark{3},
Christina P{\"o}pper\IEEEauthorrefmark{2}
}

\IEEEauthorblockA{
\IEEEauthorrefmark{1}
University of California, Berkeley\\
Berkeley, CA, USA\\
corban.villa@berkeley.edu
}

\IEEEauthorblockA{
\IEEEauthorrefmark{2}
New York University Abu Dhabi\\
Abu Dhabi, United Arab Emirates\\
\{michele.guerra, syed.khandker, christina.poepper\}@nyu.edu
}

\IEEEauthorblockA{
\IEEEauthorrefmark{3}
Northeastern University\\
Boston, MA, USA\\
\{bitsikas.e, aanjhan\}@northeastern.edu
}
}

\maketitle

\begin{abstract}
The Open Radio Access Network (O-RAN) replaces vendor-locked RANs with a modular and interoperable architecture that fosters competition and accelerates innovation. With this openness comes increased complexity and a larger attack surface, making security a critical concern. Today, assessing O-RAN security requires manually cross-referencing dozens of specifications, vendor whitepapers, and academic studies, which is error-prone and static. In this paper, we present a graph-based framework that transforms this static corpus into a single, queryable database. Our graph representation contains over 350 nodes and more than 1,250 relationships, distilled from specifications, academic papers, open-source projects, and vulnerability databases. To keep this resource current, we integrate a hybrid data extraction pipeline that couples deterministic parsing of structured specifications with Large Language Model (LLM)-assisted extraction for evolving specifications and unstructured literature. Querying the graph reveals three actionable findings within our curated corpus: critical infrastructure such as the O-DU, SMO, and O-Cloud carries dozens of specification-level threats yet has little or no empirical coverage; memory-safety weaknesses account for 11 of the 21 CWE occurrences associated with the analyzed CVEs; and fuzzing uncovered 18 of the 20 CVEs attributed to research papers. We provide the database, pipeline, and queries as open-source artifacts.
\end{abstract}


\input{Introduction}

\input{ORAN-Overview}

\input{O-RAN-Security-Graph-Database}

\input{empirical_analysis}

\input{stakeholder_insights}

\input{future_research_and_conclusion}

\bibliographystyle{IEEEtran}
\bibliography{IEEEexample}

\input{Appendix}

\end{document}

%% file: Introduction.tex
\section{Introduction}
\label{sec:Introduction}

The Open Radio Access Network (O-RAN) replaces vendor-locked, proprietary RANs with a modular, interoperable architecture that lowers costs, encourages competition, and accelerates innovation~\cite{Sandra,Motalleb2023slice}.
Since the O-RAN Alliance's inception in 2018, both industry~\cite{Nokia,Samsung,Att} and academia~\cite{Abdalla2022cntdo, Amachaghi2024survey, hablerAdversarialMachineLearning2025} have embraced this shift: ETSI~\cite{etsi} has adopted its specifications, over 30 operators have committed globally, and pilot deployments~\cite{pilot1,pilot2,NTTDocomo5GOpenRAN2021,MavenirOpenRAN2023} are emerging.

However, the disaggregated nature of O-RAN introduces increased complexity, interoperability challenges~\cite{Yungaicela2024misconfig}, and expanded attack surfaces~\cite{Singh2020Challenges,Liyanage2023security,QuadOpenRANSecurity2023}. 
O-RAN Alliance specifications define architectural threats~\cite{oran2024threat}, 3GPP standards govern the underlying cellular protocols~\cite{3gpp.33.210,3gpp.33.310,tr_38_801}, cloud providers dictate infrastructure security~\cite{awsORAN}, and orchestration engines like Kubernetes introduce distinct attack surfaces~\cite{VMwareOpenRANSecurity2021}.
Furthermore, specific deployments rely on proprietary documentation from diverse hardware and software vendors~\cite{Mavenir2021,NTTDocomo5GOpenRAN2021}. 
In response, academia and industry have produced extensive open-source frameworks~\cite{srsRAN,oaiRAN,coloran,sd-RAN,David2021NexRAN}, security investigations~\cite{Atalay2023secure,Xing24OFH,groen2024interfaces}, and surveys~\cite{amachaghiSurveyIntrusionDetection2024,Liyanage2023security,Mimran2022security,openranConciseOverview2025} with the aim to guide understanding and adoption of this evolving ecosystem.
Yet, this knowledge remains fragmented: existing research treats O-RAN elements in isolation, without providing a unified framework to understand their interdependencies.
Specifically, the research community currently lacks a comprehensive, systematic approach to analyzing and visualizing the complex relationships between (sub)components, interfaces, and threats. 

Most importantly, this fragmentation imposes a heavy cognitive burden on stakeholders. 
As a \emph{use case}, consider a network operator tasked with evaluating the security posture of the Open Distributed Unit (O-DU) for a new deployment. 
To perform a rigorous assessment today, they must manually cross-reference dozens of O-RAN Alliance specifications, vendor-specific whitepapers, and disjointed academic studies.
They must mentally map theoretical threats (to a containerized xApp, for instance) from the specifications to specific software implementations (e.g., srsRAN, O-RAN SC) and search for relevant vulnerabilities (CVEs) or available defenses. This process is manual, error-prone, and static; by the time a survey is published, the threat landscape has likely already evolved. This significantly impedes security analysts, researchers, and operators from gaining a holistic understanding of the O-RAN threat landscape.

In this paper, we present a queryable O-RAN security graph as a new paradigm for security analysis.
We introduce a graph-based analytical framework that transforms the traditional static survey into a dynamic, graph-based engine. 
Our methodology unifies theoretical threats, empirical evidence, real-world software, and known weaknesses (CVEs/CWEs) into a single ontology containing over 350 nodes and 1,250 relationships. 
Instead of traversing static text, an operator can now retrieve answers within seconds to complex questions as prioritized high-risk threats and also discover empirical defenses via a single database query.

To ensure this database serves as a living resource capable of evolving alongside the changing specifications and literature, we integrate a semi-automated extraction pipeline. 
To this end, we employ Large Language Models (LLMs) within our framework to automate ingestion of unstructured security information. 
Our evaluation demonstrates that this pipeline greatly reduces the human effort required to maintain the graph from about 15 person-hours to 65 minutes while preserving high accuracy for static content extraction.

We demonstrate the utility of our graph-based approach by identifying critical research gaps. 
For example, critical infrastructure components such as the O-DU carry dozens of specification-level threats yet have little empirical security coverage in our corpus.
Our analysis also finds that memory-safety weaknesses account for 11 of the 21 CWE occurrences associated with the analyzed CVEs, motivating greater use of memory-safe languages. Moreover, fuzzing uncovered 18 of the 20 CVEs in our dataset that were attributed to research papers, highlighting its value as a hardening technique for the ecosystem.

In summary, the main contributions of this paper are:
\begin{compactenum}
    \item \textbf{A Graph-Based Methodology.} We propose a novel framework to systematize O-RAN security, moving beyond static text to a queryable, relational database that unifies specifications, literature, software implementations, and vulnerability databases into a single ontology of over 350 nodes and 1,250 relationships.
    \item \textbf{LLM-Integrated Automation Pipeline.} We design and evaluate a semi-automated pipeline using LLMs to extract structured security data from heterogeneous sources, reducing curation effort from $\sim$15 person-hours to 65 minutes while ensuring the database remains scalable and up-to-date.
    \item \textbf{Stakeholder-Specific Insights.} We leverage the graph to uncover non-obvious relationships and gaps, providing actionable recommendations for researchers, operators, agencies, and developers---including the critical under-study of the O-DU, %
    memory safety as the dominant weakness class in our CWE analysis (11 of 21 occurrences), and fuzzing as the source of 18 of the 20 CVEs attributed to research papers in our dataset.
\end{compactenum}

\noindent\textbf{Artifact availability:} The artifact\footnote{https://github.com/MicheleGuerra/SOK-Oran-Security} include the graph-based O-RAN security model, an interactive demo, full codebase, Cypher\footnote{Cypher is Neo4j's declarative graph query language.} query set, GUI, prompts, curated CSVs, and the instructions referenced in Appendix~\ref{appendix:database-access}.

%% file: ORAN-Overview.tex
\section{O-RAN Overview}
\label{sec:O-RAN-Overview}

\begin{figure}[t]
\centering
\includegraphics[width=.8\columnwidth]{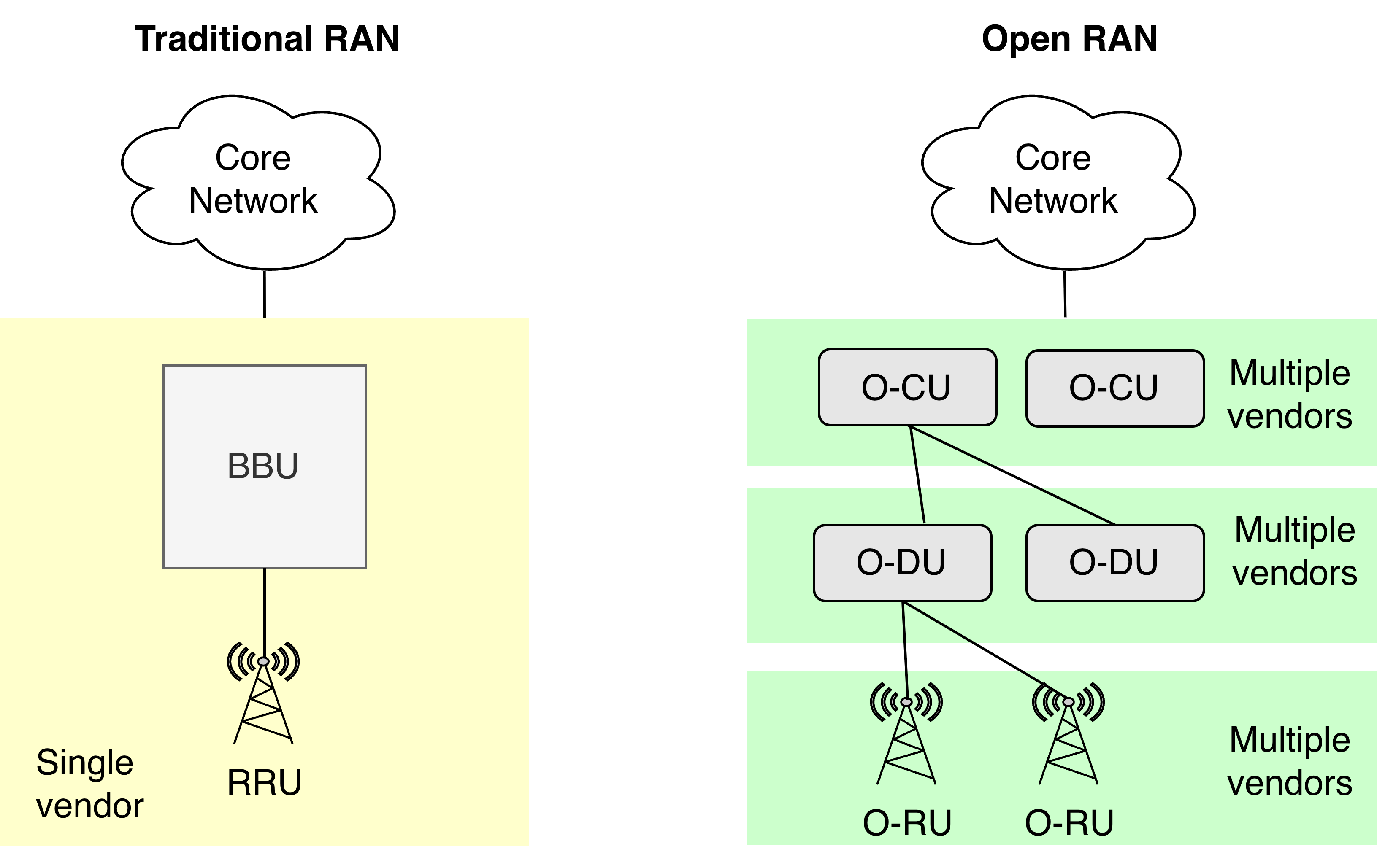}
\caption{Traditional RAN vs.~O-RAN Architectures.}
\label{fig:traditional}
\end{figure}

Figure~\ref{fig:traditional} illustrates the architectural differences between traditional RAN and O-RAN. Decoupling hardware and software at various levels enables a vendor-neutral ecosystem, fostering collaboration and competition among equipment suppliers and software developers.

\subsection{O-RAN Components and Interfaces}
\label{subsec:component}

According to the specification~\cite{oran2024architecture}, the key O-RAN \emph{components} are as follows (Fig.~\ref{fig:O-RAN_architecture}):

\begin{figure}[tb]
\centering
\includegraphics[width=0.8\columnwidth]{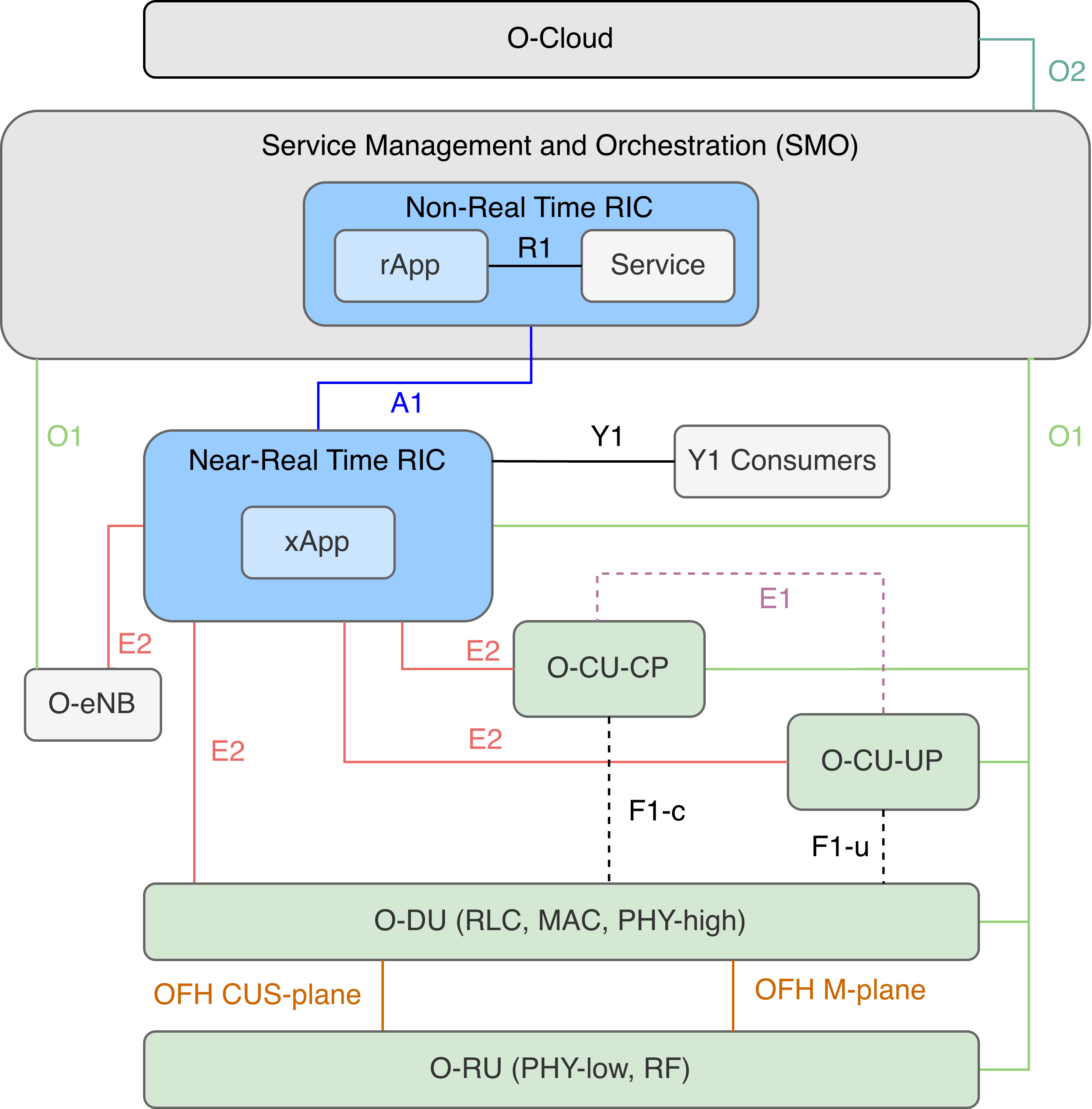}
\caption{O-RAN Components and Interfaces.}
\label{fig:O-RAN_architecture}
\end{figure}

\noindent\textbf{SMO:} The Service Management and Orchestration (SMO) manages network functions, incl.~resource allocation, configuration, fault management, and performance monitoring. 

\noindent\textbf{Near-RT RIC:} Enables near real-time network control and optimization through xApp, which operates with latency typically within 10 to 1,000 milliseconds.

\noindent\textbf{O-CU:} The Open Central Unit Control Plane (O-CU-CP) and User Plane (O-CU-UP). The former executes functions incl.~sessions, mobility management, and the RRC protocol for UE radio connections. It communicates with O-DU via the F1-C interface. The latter manages data traffic between the network and the UE. Separation from the O-CU-CP allows it to focus on data transfer while minimizing latency.

\noindent\textbf{O-DU:} The Open Distributed Unit (O-DU) handles real-time radio signal processing and manages lower radio protocol layers (e.g., MAC, RLC). It coordinates with O-RU and O-CU for radio resource scheduling and signal en-/decoding, ensuring efficient data transmission, optimal spectrum use, and low-latency communication.

\noindent\textbf{O-RU:} The Open Radio Unit (O-RU) manages the transmission and reception of radio signals between UEs and the network. It converts  analog radio signals to digital data streams and handles tasks like RF processing, Fast Fourier Transform, beam-forming, etc.%

\noindent\textbf{O-eNB:} To support LTE, O-RAN employs Open evolved Node B (O-eNB), which manages user and control plane functions. It handles radio signal processing and protocol stacks, similar to traditional eNBs, but is designed to be open and interoperable with equipment from other vendors, thus enhancing flexibility in 4G networks.

\noindent\textbf{O-Cloud:} The infrastructure that hosts various O-RAN functions (e.g., Near-RT RIC, O-CU, and O-DU) and the supporting software components (e.g., operating system, containers). It provides the computational, networking, and storage resources necessary for efficient network operations.

\noindent\textbf{xApp} and \textbf{rApp:} xApps and rApps are hosted on the Near and Non-Real-Time \textbf{(Non-RT) RIC}s, respectively.
\emph{xApps} enable real-time tasks, incl.~traffic steering, load balancing, geo-location, and interference classification. 
\emph{rApps} offer long-term benefits, incl.~energy savings, quality optimization, and predictive maintenance, while operating with higher latency. \emph{KPMs} (Key Performance Measurements) enable continuous monitoring of performance, incl.~resource utilization, throughput, latency, and error rates.

\vspace{1em}

\noindent The key O-RAN \emph{interfaces} are: 

\noindent\textbf{A1:} The A1 interface connects the Non-RT RIC to the Near-RT RIC, supporting policy management, enrichment information, and ML model management services.

\noindent\textbf{O1:} The O1 interface connects all managed elements for the SMO. It supports FCAPS (Fault, Configuration, Accounting, Performance,  Security) and management functions including discovery, registration, configuration, and monitoring.

\noindent\textbf{O2:} The O2 interface connects the SMO and O-Cloud, enabling management of O-Cloud infrastructure and network functions.%

\noindent\textbf{E2:} The E2 interface connects the Near-RT RIC with E2 nodes, supporting performance monitoring through KPM channels and passing control commands to the O-CU.

\noindent\textbf{OFH:} Communication between O-DU and O-RU occurs via the Open Fronthaul (OFH) Interface, consisting of four planes: Control (C), User (U), Synchronization (S), and Management (M). Planes handle management and user data transmission, clock synchronization, and O-RU configurations, respectively.%

\noindent\textbf{Y1:} The Y1 interface connects the Near-RT RIC to Y1 consumers, exposing policy management entities, network management functions, and AI/ML model management.

\noindent\textbf{R1:} The R1 interface between the rApps and the Non-RT RIC allows rApps to access data analytics, policy information, and ML models provided by the Non-RT RIC.

\subsection{O-RAN Security Ecosystem}

The O-RAN security landscape is defined by a complex interplay between standardization bodies, open-source communities, and external regulators. 
The O-RAN Alliance spearheads standardization primarily through specifications developed by 11 technical working groups and 2 focus groups, each addressing distinct aspects of the O-RAN architecture~\cite{spec}. Notably, the security focus group was upgraded in 2022 to a full working group (WG11) to address rising concerns~\cite{ORANFocusGroupToWGBlog}. WG11 publishes the foundational O-RAN Threat Model~\cite{oran2024threat}, which catalogs risks across the SMO, RICs, and open interfaces, and advocates for a Zero Trust Architecture (ZTA)~\cite{oran_zta,zeroTrustORAN2025}. Complementing these specifications, the O-RAN Software Community (OSC) provides reference implementations~\cite{osc,ORANOSCLaunch2019}, though these emerging software stacks have recently begun to exhibit their own vulnerabilities, evidenced by the disclosure of the first O-RAN specific CVEs in late 2023~\cite{hungAnomalyDetectionMitigating2025}.

Beyond the Alliance, the ecosystem is heavily influenced by industry and government scrutiny. Major vendors and operators have published whitepapers detailing secure implementation strategies, often focusing on cloud security and micro-segmentation~\cite{dellORANSecurity2023, NTTDocomo5GOpenRAN2021, VMwareOpenRANSecurity2021, Mavenir2021, Ericsson2021, fujitsu}. Conversely, government bodies, including the German BSI~\cite{OpenRANRiskAnalysis2022}, the EU~\cite{NIS2022}, and the Quad Critical Technology Group~\cite{QuadOpenRANSecurity2023}, have issued critical reports highlighting risks inherent to multi-vendor environments. These reports emphasize expanded attack surfaces and the lack of concrete mitigation strategies for AI/ML components~\cite{CISARadioAccessNetworkSecurity, CSRICVIIIOpenRAN2022, VellietOpenRAN2022}.

\begin{figure}[tb]
    \includegraphics[width=1.14\columnwidth,trim={0.95in 0.1in 0.0in 1.5in},clip]{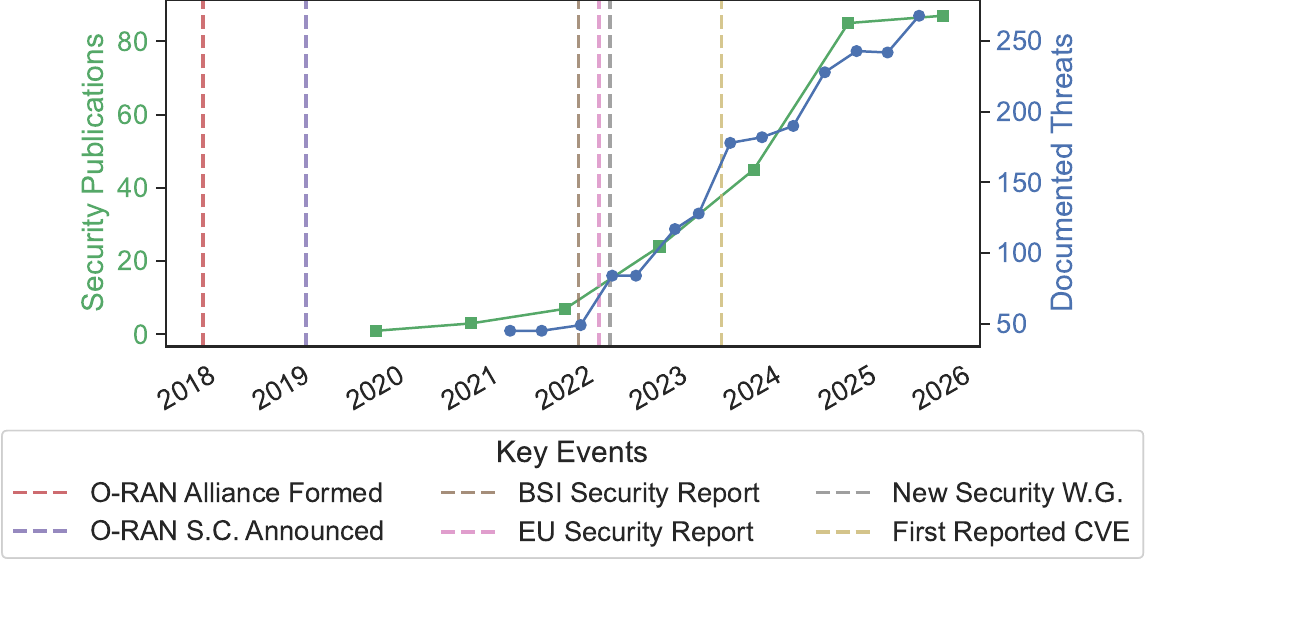}
    \caption{O-RAN Security Timeline: Publications, Threats, Key Events. The green and blue curves depict the number of security publications and documented threats, respectively.}
    \label{fig:security-timeline}
\end{figure}

Despite this wealth of documentation, a unified view remains elusive. As shown in Figure~\ref{fig:security-timeline}, while the number of documented threats in the specifications has grown steadily (blue line), empirical security research (green line) has only recently surged to match it~\cite{singhInsightsTrendsOpen2025}. Current efforts remain fragmented: specifications define theoretical risks, industry reports focus on deployment best practices, and academia explores specific attack vectors, such as data poisoning or intrusion detection, in isolation~\cite{Liyanage2023security,Mimran2022security,rattiEnhancingMobileNetwork2024,tsourdinisAIDrivenNetworkIntrusion2024,Abdalla2022cntdo,motallebSecureIntelligentORAN2024}. This disconnect emphasizes the need for a systematic approach to thoroughly distill the O-RAN security landscape.

%% file: O-RAN-Security-Graph-Database.tex
\section{O-RAN Security Graph Database}
\label{sec:graph-database}

In this section, we introduce our novel graph-based approach to systematize O-RAN security across complex threat modeling specifications~\cite{oran2024threat}, government reports~\cite{OpenRANRiskAnalysis2022, CSRICVIIIOpenRAN2022, NIS2022, CISARadioAccessNetworkSecurity}, empirical academic work, software implementations, and vulnerabilities.
We empirically analyze it in Section~\ref{sec:empirical-analysis} and practically demonstrate the utility of this graph representation in Section~\ref{sec:insights}, where we elicit specific takeaways for O-RAN stakeholders---leveraging queries that effectively span our source documents.

%

\begin{comment}

\begin{figure}[t]
  \centering
  \includegraphics[width=1\linewidth]{figures/general/process-diagram2.pdf}
  \caption{\textbf{Systematizing O-RAN Security with Graphs:} 
  %
  \textbf{Sec.~\ref{sec:graph-database}} describes the graph construction methodology, including security sources and automation.
  %
  \textbf{Sec.~\ref{sec:insights}} elicits insights for various stakeholders, demonstrating the graph's novel versatility with queries.}
  \Description{}
  \label{fig:process-diagram}
\end{figure}

\end{comment}

\begin{figure*}[t]
  \centering
  \includegraphics[width=.93\textwidth,trim={0.in 0.1in 0.in 0.2in},clip]{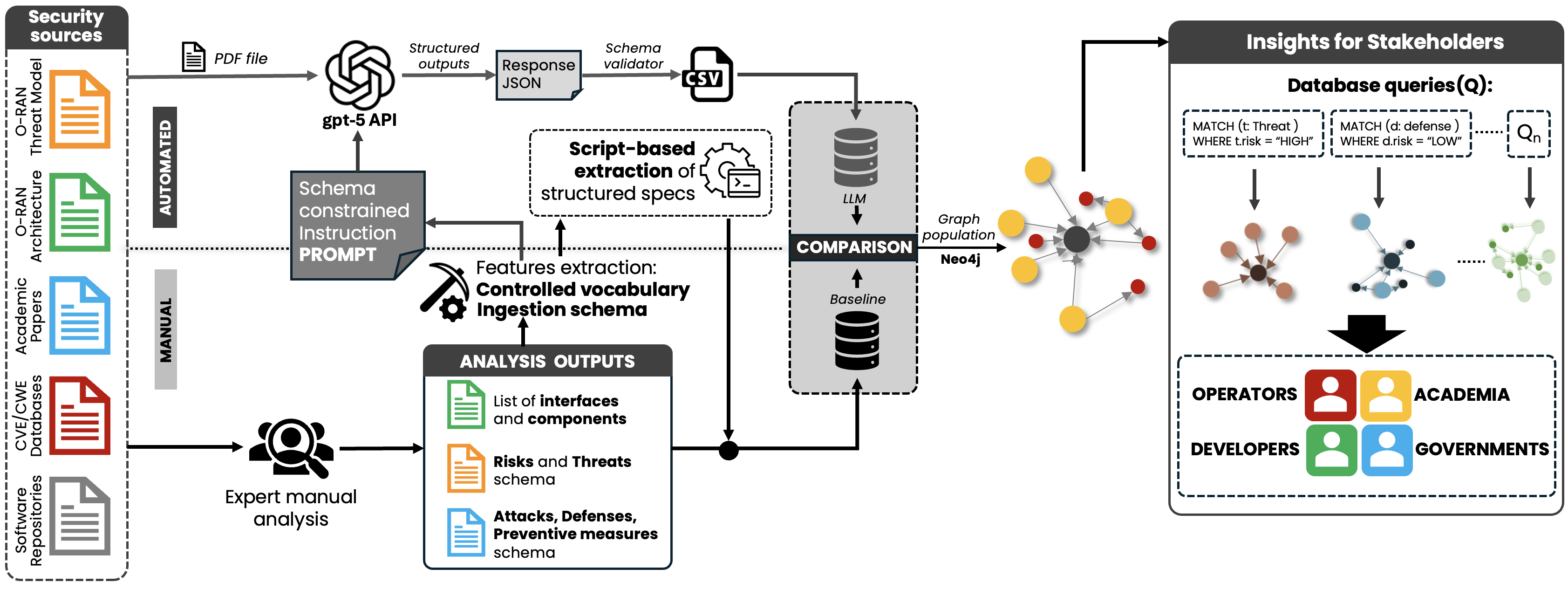}
  \caption{Overview of our Approach and Built Pipeline.}
  \label{fig:pipeline}
\end{figure*}

\subsection{Rationale}

Prior works have attempted to distill the challenging O-RAN security landscape through comprehensive surveys~\cite{poleseUnderstandingORANArchitecture2023,Liyanage2023security,Mimran2022security,soltaniIntelligentControl6G2025,hablerAdversarialMachineLearning2025,rebecchiRevealingThreatLandscape2024,amachaghiSurveyIntrusionDetection2024,klementSecuring6GTransition2024,tabibanSignalingStormORAN2024}. While valuable, these contributions are fundamentally limited by their \textit{static presentation}. In a traditional survey, components and threats are enumerated in linear lists or static tables. This format hinders the logical connections between entities; it is infeasible to query a PDF to trace the ``blast radius'' of a compromised interface or instantly identify which software implementations are affected by a newly discovered theoretical threat. Moreover, static surveys inevitably become obsolete as specifications are updated, requiring new publications to reflect the state of the art.

We propose a paradigm shift from static documentation to a dynamic graph database, which is uniquely suited to O-RAN security by naturally modeling the network's interconnected topology and unifying heterogeneous data sources into a single ontology. However, constructing such a graph manually is labor-intensive and prone to scalability issues given the verbose and specifications' ever-evolving nature. To address this, we integrate LLMs into our pipeline. By leveraging LLMs to parse unstructured text and map it to our graph schema, we bridge the gap between massive, disjointed documentation and structured, queryable insights, ensuring the system remains a living, up-to-date resource.

\emph{Comparison with Existing Graph Approaches.} 
Graph-based security analysis has been explored in other domains~\cite{granataSystematicAnalysisAutomated2024}. \etal{Böhm}{\cite{bohmGraphbasedVisualAnalytics2018}} utilized STIX for incident visualization, while others have modeled privacy anti-patterns~\cite{kunzPrivacyPropertyGraph2023, munirPURLSafeEffective2024}. Regarding automated extraction, \etal{Zhao}{\cite{zhaoCyberThreatIntelligence2020}} leveraged graph convolutional networks to extract threat intelligence from unstructured text, while \etal{Branescu}{\cite{branescuAutomatedMappingCommon2024}} used transformers to map CVEs to MITRE ATT\&CK.
In the cellular domain, \etal{Pacheco}{\cite{pachecoAutomatedAttackSynthesis2022}} and \etal{Ishtiaq}{\cite{Hermes2024}} utilized NLP to extract Finite State Machines (FSMs) from protocol specifications to discover design issues. Within O-RAN specifically, graphs have been used to model xApp conflicts~\cite{zolghadrLearningReconstructingConflicts2025}, optimize graph neural network training~\cite{balakrishnanEnhancingORANSecurity2024}, represent Near-RT RIC topology~\cite{orhanConnectionManagementXAPP2021}, and construct specific attack knowledge graphs for 6G~\cite{wangCyberAttackBehaviorKnowledge2022}.

Our approach differs in scope and integration. Instead of focusing on a \textit{single context} (e.g., protocol FSMs, xApp conflicts, specific attack vectors), our work acts as a \textit{meta-analysis tool}. We unify the disjointed layers of the O-RAN ecosystem into a common analytical representation. By augmenting this unification with LLM-driven extraction, we enable ontological insights that are difficult to obtain when the sources are analyzed in isolation.

\subsection{Our Methodology}
\label{sec:merged-methodology}

We constructed the O-RAN security graph by aggregating information across disjoint contexts---theoretical threat models, empirical academic research, and hardened software implementations. Our methodology utilizes a \emph{hybrid extraction pipeline} (Fig.~\ref{fig:pipeline}) that couples deterministic parsing for structured specifications with an LLM-assisted module for unstructured literature, anchored by human-in-the-loop supervision.

\subsubsection{Data Sources and Ontology}
We defined a graph ontology consisting of \emph{Components}, \emph{Interfaces}, \emph{Threats}, \emph{Attacks}, \emph{Defenses}, and \emph{Software}. The resulting database contains over 350 nodes and more than 1{,}250 relationships, populated from five primary sources:

\begin{enumerate}[leftmargin=*, itemsep=0pt, parsep=0pt, topsep=0pt]
    \item \textbf{Architectural Specifications:} Canonical components and interfaces derived from O-RAN Alliance specifications~\cite{oran2024architecture}.
    \item \textbf{Threat Models:} Theoretical threats and risk scores extracted from the O-RAN Threat Model~\cite{oran2024threat}.
    \item \textbf{Empirical Literature:} A curated set of attacks, defenses, and preventive measures (e.g., fuzzing~\cite{Yang2024osc,dessources5GORANSecurity2024,hungAnomalyDetectionMitigating2025}) from top systems/security venues and keyword-based searches on Google Scholar \& Semantic Scholar. We include only works demonstrating proof-of-concepts or experimental validation. The full current list of academic works is provided in Table~\ref{tab:threats}.
    \item \textbf{Software Implementations:} Open-source projects (e.g., srsRAN, O-RAN SC) mapped to the components they implement and reference in literature.
    \item \textbf{Vulnerability Databases:} Common Vulnerabilities and Exposures (CVEs) and Weaknesses (CWEs) from MITRE associated with the identified software (in some cases incl.~further details from \cite{Yang2024osc}).
\end{enumerate}

\begin{table}[tb]
\centering
\caption{Graph ontology: node types and relationships.}
\label{tab:ontology}
\footnotesize
\setlength{\tabcolsep}{3pt}
\renewcommand{\arraystretch}{1.05}
\begin{tabularx}{\linewidth}{@{}l X@{}}
\toprule
\textbf{Node Type} & \textbf{Description} \\
\midrule
Component       & Architectural elements (e.g., O-DU, Near-RT RIC) \\
Interface       & Standardized interfaces (e.g., E2, A1) \\
Threat          & Specification-level risks with severity scores \\
Attack          & Empirical exploits with proof-of-concept \\
Defense         & Runtime protective mechanisms (e.g., anomaly detection) \\
Preventive Measure  & Pre-deployment hardening (e.g., fuzzing, testing) \\
Software        & Open-source implementations (e.g., srsRAN) \\
CVE             & Known vulnerabilities from MITRE \\
CWE             & Weakness classifications from MITRE \\
\midrule
\textbf{Relationship} & \textbf{Direction} \\
\midrule
\textsc{targets}         & Threat/Attack $\rightarrow$ Component/Interface \\
\textsc{secures}         & Defense/Preventive Measure $\rightarrow$ Component/Interface \\
\textsc{implements}      & Software $\rightarrow$ Component \\
\textsc{affects}         & CVE $\rightarrow$ Software \\
\textsc{associated\_with} & CVE $\rightarrow$ CWE \\
\bottomrule
\end{tabularx}
\end{table}

Table~\ref{tab:ontology} summarizes the graph ontology. Entities and their relationships are loaded into a Neo4j~\cite{neo4j} graph database, chosen for its expressive query language and self-hosting capabilities.  

\subsubsection{Hybrid Extraction Pipeline}
\label{subsec:automation-role}
To ensure accuracy and scalability, we employ two complementary extraction mechanisms.

\emph{Deterministic Parsing for Structured Specs.}
For documents with consistent tabular layouts, such as the O-RAN Threat Model~\cite{oran2024threat}, we utilize deterministic Python scripts to automate manual extraction. Tables are exported to CSVs and processed via regular expressions. In our workflow, this method reduced the extraction time for the Threat Model from \emph{5 hours} (manual) to under \emph{2 minutes} (script-based). We found this method transferable to similarly formatted documents~\cite{NIS2022,euCybersec5g}.
The design of this pipeline was informed by a phase of manual analysis. We carefully examined the specification tables to identify canonical entities, recurring structural patterns, implicit relationships, and textual inconsistencies. This process not only enabled us to define normalization rules and controlled vocabularies, but also surfaced eleven typographical errors (App.~\ref{appendix:typos}) in the underlying specifications. These inconsistencies were later automatically detected when the database rejected relationships referencing undefined assets (Table~\ref{table:typos}).
The insights gained from this manual inspection directly shaped the deterministic parser and the schema constraints later used in the LLM-based extraction pipeline (Section~\ref{subsec:update-database}).

\emph{LLM-Assisted Extraction for Unstructured Text.}
For heterogeneous academic papers where deterministic parsing fails, we integrate a schema-constrained LLM module. We call \texttt{GPT-5}\footnote{OpenAI, \emph{GPT-5}, \url{https://platform.openai.com/docs/models/gpt-5}} via the API using \texttt{reasoning.effort=medium}, \texttt{verbosity=low}, and an output budget of 8k tokens.
We adopt a schema-first prompting strategy: the prompt enumerates valid column headers and forces the model to map entities to our controlled vocabulary (derived from~\cite{oran2024architecture}). All outputs use \emph{Structured model outputs}\footnote{OpenAI, \emph{structured model outputs}, \url{https://platform.openai.com/docs/guides/structured-outputs}} to validate JSON before CSV conversion. Prompts were iteratively refined and then regularized with the OpenAI Optimizer\footnote{OpenAI, \emph{Prompt optimizer}, 
\url{https://platform.openai.com/chat/edit?models=gpt-5&optimize=true}} to minimize format drift.
This LLM automation significantly accelerates synthesis: processing our corpus of 35 papers required \emph{65 minutes} via the LLM pipeline, compared to approximately \emph{15 person-hours} for manual curation (avg.~25 min/paper).
We leveraged the \texttt{neo4j}, \texttt{pandas}, and \texttt{openai} libraries to construct our end-to-end system in 1728 lines of Python code. 
All prompting templates are made available. %

\subsubsection{Maintenance and Reproducibility}
\label{subsec:update-database}

\begin{figure}[tb]
  \centering
  \includegraphics[width=0.85\linewidth]{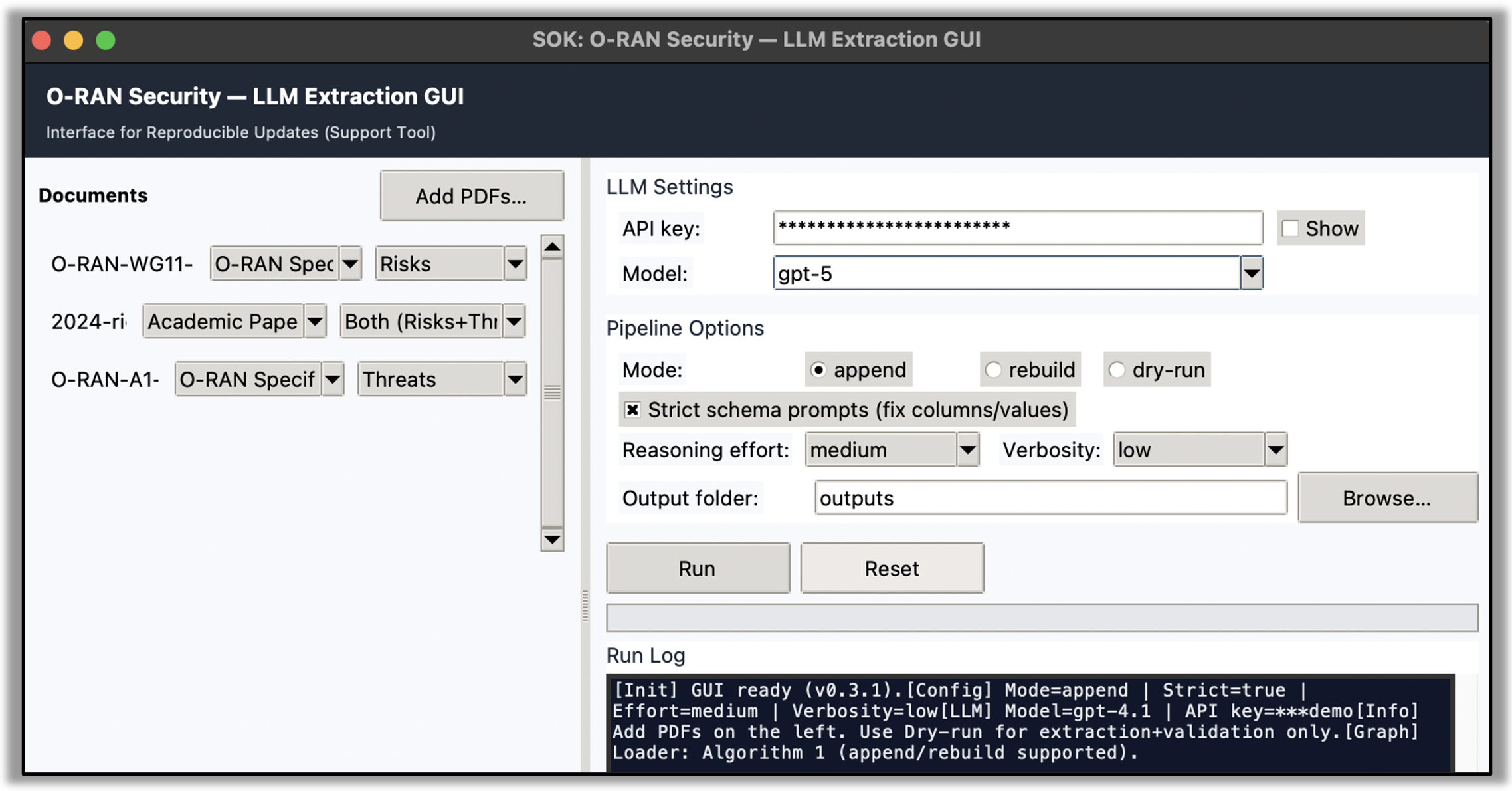}
  \caption{GUI for reproducible pipeline updates.}
  \label{fig:GUI}
\end{figure}

To ensure the graph remains a living resource, we developed a GUI (Fig.~\ref{fig:GUI}) that orchestrates the entire pipeline. The interface allows users to upload new PDFs, select the document type (academic paper or specification), and execute graph updates. Runs can be executed in \emph{append} mode to extend the graph or \emph{rebuild} mode to recreate it. Run logs are versioned to enable exact reproducibility. Updating threats upon new specification releases took less than one hour manually, or approx.~\emph{7 minutes} using our guided automated pipeline (excl.~LLM processing time). We validated the extraction on earlier revisions of the O-RAN specifications whose table organization and field names differ slightly from the current release. Using the strict prompt variant (restating headers and admissible values) and the schema validator, the module yielded structured outputs consistent with our graph schema, requiring no changes to the ingestion code.

\begin{figure}[tb]
  \centering
  \includegraphics[width=1.\linewidth]{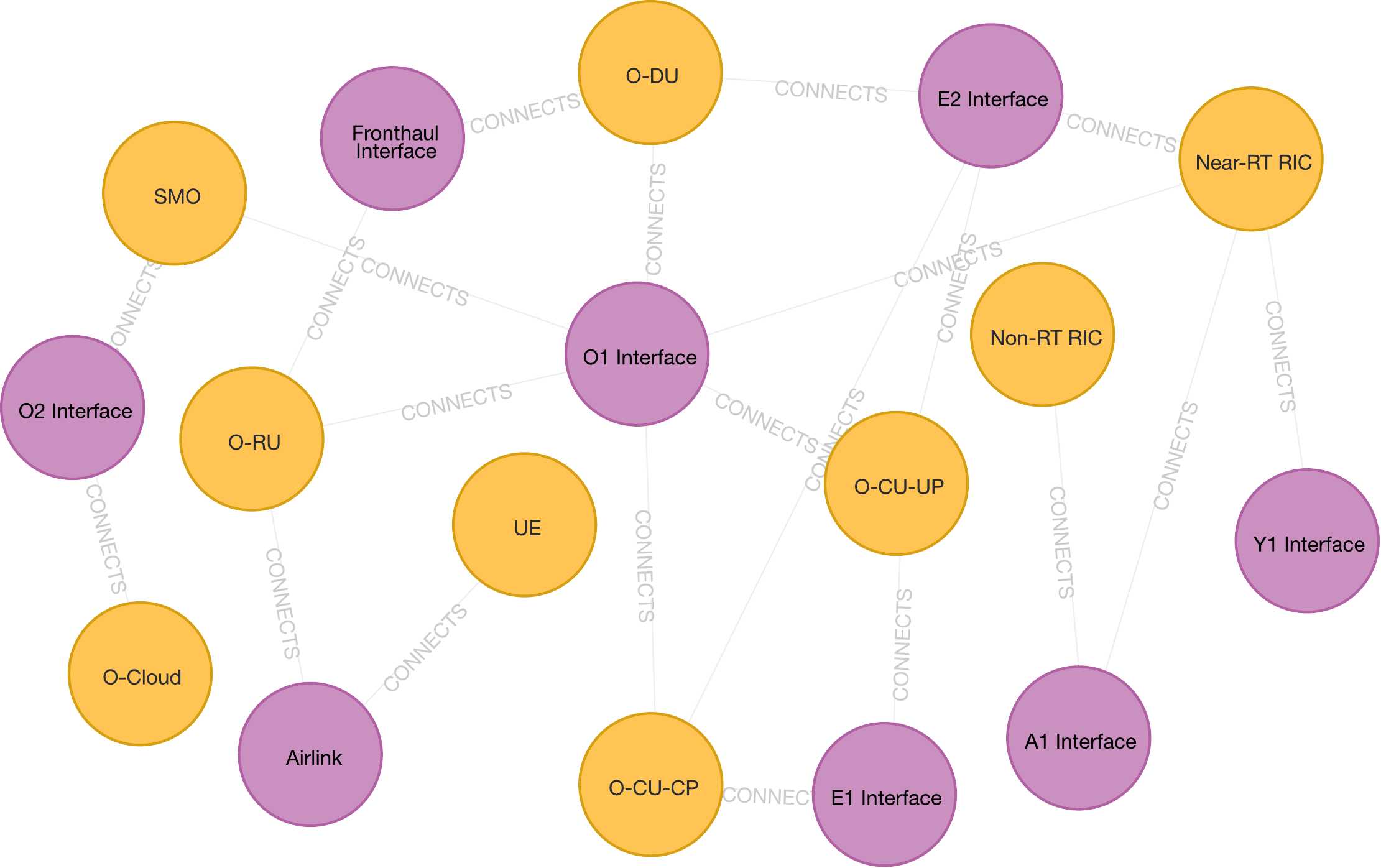}
  \caption{Graph-based O-RAN Topology: O-RAN components (\codot \, yellow) and interfaces (\idot \, violet). Consistency with specifications~\cite{oran2024architecture} provides a cross-validation of our dataset. The centrality of the O1 interface becomes apparent, as does the E2 interface; all other interfaces connect two components. %
  }
  \label{fig:components}
\end{figure}

\subsubsection{A Queryable Security Database}
A crucial benefit of this approach is the shift from static documents to dynamic exploration. Users can leverage the graph visualization (Fig.~\ref{fig:components}) alongside expressive queries (List.~\ref{listing:samplequery}) to elicit nuanced security insights iteratively. Beyond manual query formulation, we further extend this paradigm by enabling natural language interaction with the database through LLMs. Instead of crafting Cypher queries directly, users may describe high-level security questions in natural language, which are translated into structured graph queries aligned with the database schema, lowering the barrier for non-technical users. Implementation details, prompting templates, and deployment instructions are included in the accompanying artifact.

In practice, the workflow combines query synthesis and exploratory reasoning under a schema-constrained prompting strategy. An agentic architecture allows the LLM to iteratively generate queries, execute them via a local Neo4j service, and refine results, with a temporary \texttt{ngrok}-based endpoint enabling remote database access during experimentation. Insights and queries are reviewed by users to detect and mitigate unsupported outputs.

\begin{figure}[t]
\begin{minipage}{\linewidth}
\begin{lstlisting}[style=CStyle, frame=single, caption={Under-scrutinized O-RAN Components: Return components with no empirical attacks, defenses, preventive measures, or CVEs.},label=listing:samplequery]
MATCH (c: Component)
WHERE NOT ((c)<-[:TARGETS]-(:Attack))
  AND NOT ((c)<-[:SECURES]-(:Defense))
  AND NOT ((c)<-[:SECURES]-(:PreventiveMeasure))
OPTIONAL MATCH (s:Software)-[:IMPLEMENTS]->(c)
WHERE NOT ((s)<-[:AFFECTS]-(:CVE))
RETURN c.name AS component, collect(s.name) AS software
\end{lstlisting}
\end{minipage}
\end{figure}

%% file: empirical_analysis.tex
\section{Empirical O-RAN Security Analysis}
\label{sec:empirical-analysis}

\begin{figure*}[t]
  \centering
  \includegraphics[width=\linewidth]{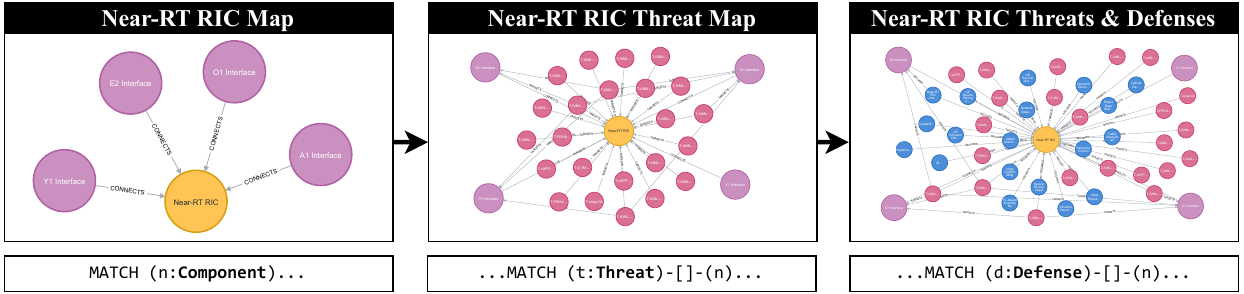}
  \caption{Novelty of Queryable Threat Mapping: These incremental threat model investigation examples for the Near-RT RIC underscore the vital paradigm shift provided by a queryable threat model---as opposed to traditional representations.}
  \label{fig:incremental-query}
\end{figure*}

In this section, we leverage our graph database to visualize and analyze the current state of empirical O-RAN security research. By mapping  disparate studies onto the architectural topology (Fig.~\ref{fig:experimental-attack} and \ref{fig:mitigation-mapping}), we identify distinct concentrations of offensive and defensive efforts, as well as significant coverage gaps. 

\subsection{Empirical Attack and Defense Landscape}
\label{subsec:empirical-landscape}

We organize the empirical literature into three architectural domains: intelligent control (RICs and xApps), which dominates current research; open interfaces, which expose new interception points; and infrastructure (O-DU, O-CU), which remains significantly under-studied.

\emph{Attacks on Intelligent Control.}
Adversarial attacks against Machine Learning (ML) models are prevalent. Chiejina et al.~\cite{Chiejina2024MLattack} demonstrated how malicious xApps can poison spectrogram and KPM data in the Near-RT RIC to degrade network performance. Similarly, malicious xApps have been used to detect rogue base stations~\cite{Huang2023RBS} or classify cellular attacks in real-time~\cite{Wen2024Spector}. Vulnerabilities within the Near-RT RIC itself allow for critical disruptions; the "Bearer Migration Poisoning" (BMP) attack~\cite{Sanaz2023BPM} exploits the RIC to divert user plane traffic into routing blackholes. Other works have explored logic conflicts and isolation compromises~\cite{awadXAppsDDoSAttacks2024,jiangOZTrustORANZeroTrust2023,kouchakiAdvancingORANSecurity2024}.

\emph{Defenses for Intelligence.}
The majority of defensive proposals target this layer. Proposed mechanisms include the \textit{xApp Repository Function (XRF)} for scalable token-based authentication~\cite{Atalay2023secure} and Zero-Trust Architectures (ZTA) that enforce access control via dynamic tracing~\cite{jiangOZTrustORANZeroTrust2023}. Anomaly detection is another major focus; researchers have proposed using xApps as security sensors to detect jamming~\cite{rumeshFederatedLearningAnomaly2024} or classify network intrusions based on telemetry~\cite{groenTRACTORTrafficAnalysis2023, Wen2024Spector}. Distillation techniques have also been proposed to transfer knowledge from "teacher" models to secure student models against poisoning~\cite{Chiejina2024MLattack}.
Figure~\ref{fig:incremental-query} provides a representative example of iterative database queries and graph representations for the Near-RT RIC component, its known threats, and defense mechanisms.

\emph{E2 Interface.}
The E2 interface is a critical attack vector. Demonstrated exploits include Denial of Service (DoS) via crafted E2Manager requests and protocol state manipulation between E2 nodes~\cite{Hung2024E2, kimSimulationARPSpoofing2024, endlessSubscriptionsEuroSP2025}. Preventing these attacks requires rigorous input validation; recent work has introduced fuzzing frameworks specifically for the E2 interface to uncover these bugs~\cite{Yang2024osc}. \etal{Groen}{\cite{groen2024interfaces}} note that while IPsec can secure E2, it adds significant overhead (57 bytes per packet plus encryption latency), potentially impacting real-time performance~\cite{Groen2023cost}.

\emph{A1 Interface.}
The A1 interface has been shown to be vulnerable to Man-in-the-Middle (MitM) attacks using ARP poisoning, allowing attackers to leak data or manipulate policy traffic if TLS is not strictly enforced~\cite{Tiberti2022mitma}. Thimmaraju et al.~\cite{Thimmaraju2024a1interface} developed a testing tool that revealed TLS and OAuth are often unsupported on policy servers (e.g., $\mu$ONOS, OSC), leaving them exposed to such attacks.

\emph{Open Fronthaul (OFH)}
The physical transport layer presents critical risks. Researchers have developed tools to launch C-Plane DoS attacks via spoofed MAC addresses~\cite{Liao2022dos} and demonstrated exploits leveraging untrusted vendors to degrade performance without physical access~\cite{Xing24OFH}. While MACsec is frequently recommended as a solution~\cite{Cho2021openfront, Dik2021FH}, its deployment is often hindered by performance concerns~\cite{groen2024interfaces}.

\emph{Infrastructure (O-DU, O-CU).}
In contrast to the RICs and interfaces, the lower-layer infrastructure remains significantly under-studied.
\etal{Groen}{\cite{groenTIMESAFETimingInterruption2024}} provide one of the few works targeting the O-DU and O-RU, focusing on timing interruption attacks. Similarly, only a single work in our dataset scrutinizes the O-DU, O-CU, and F1 interfaces comprehensively~\cite{dessources5GORANSecurity2024}. This gap is critical, as the O-DU processes real-time radio signals; vital for network availability.

\subsection{Identified Coverage and Gaps}
\label{subsec:summary}

\begin{figure}[tb]
  \centering
  \includegraphics[width=1.\linewidth]{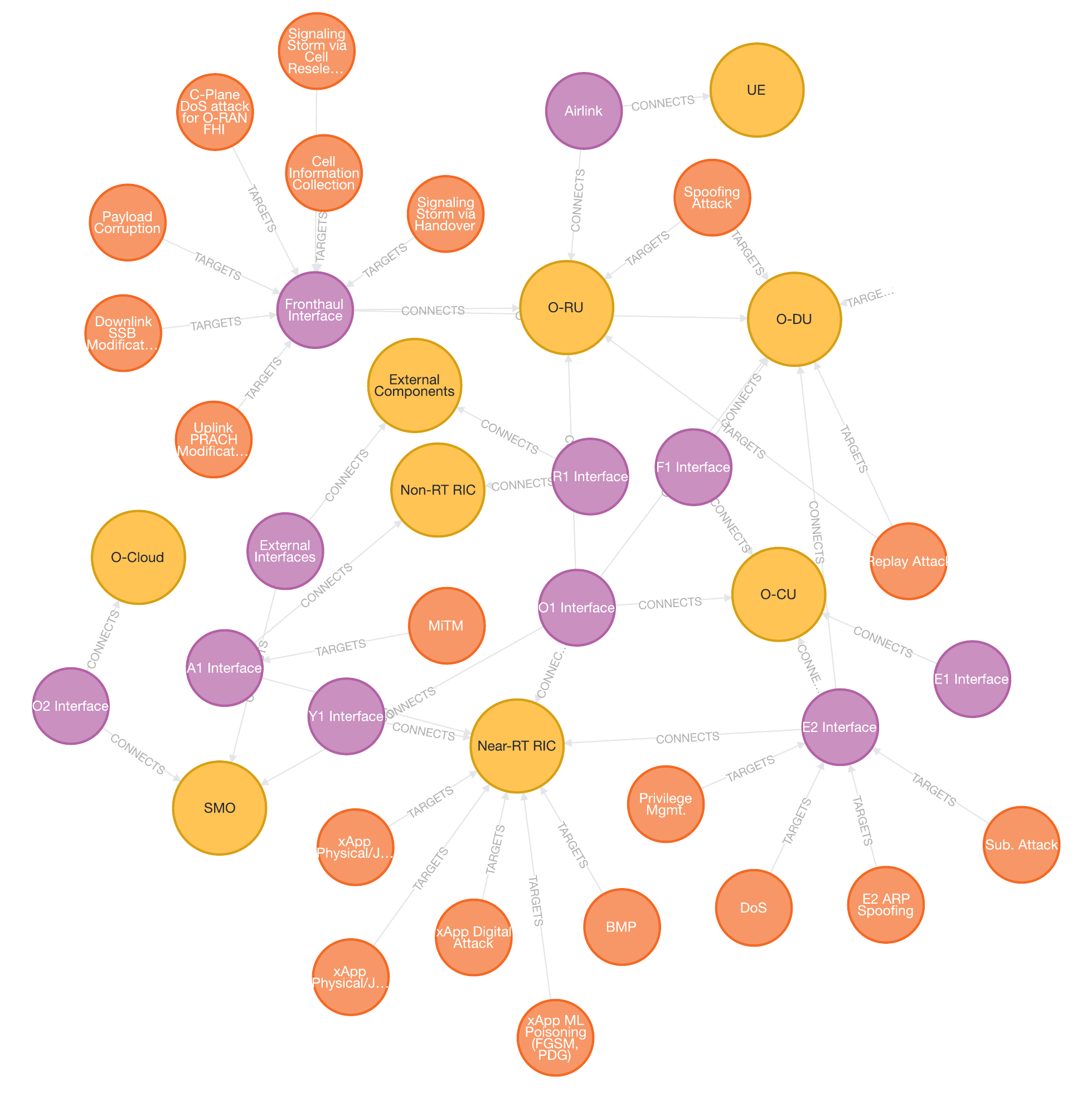}
  \caption{Map of O-RAN Attacks. The threats (\ordot) are linked to components (\codot) or interfaces (\idot).}
  \label{fig:experimental-attack}
\end{figure}

\begin{figure}[tb]
  \centering
  \includegraphics[width=1.\linewidth]{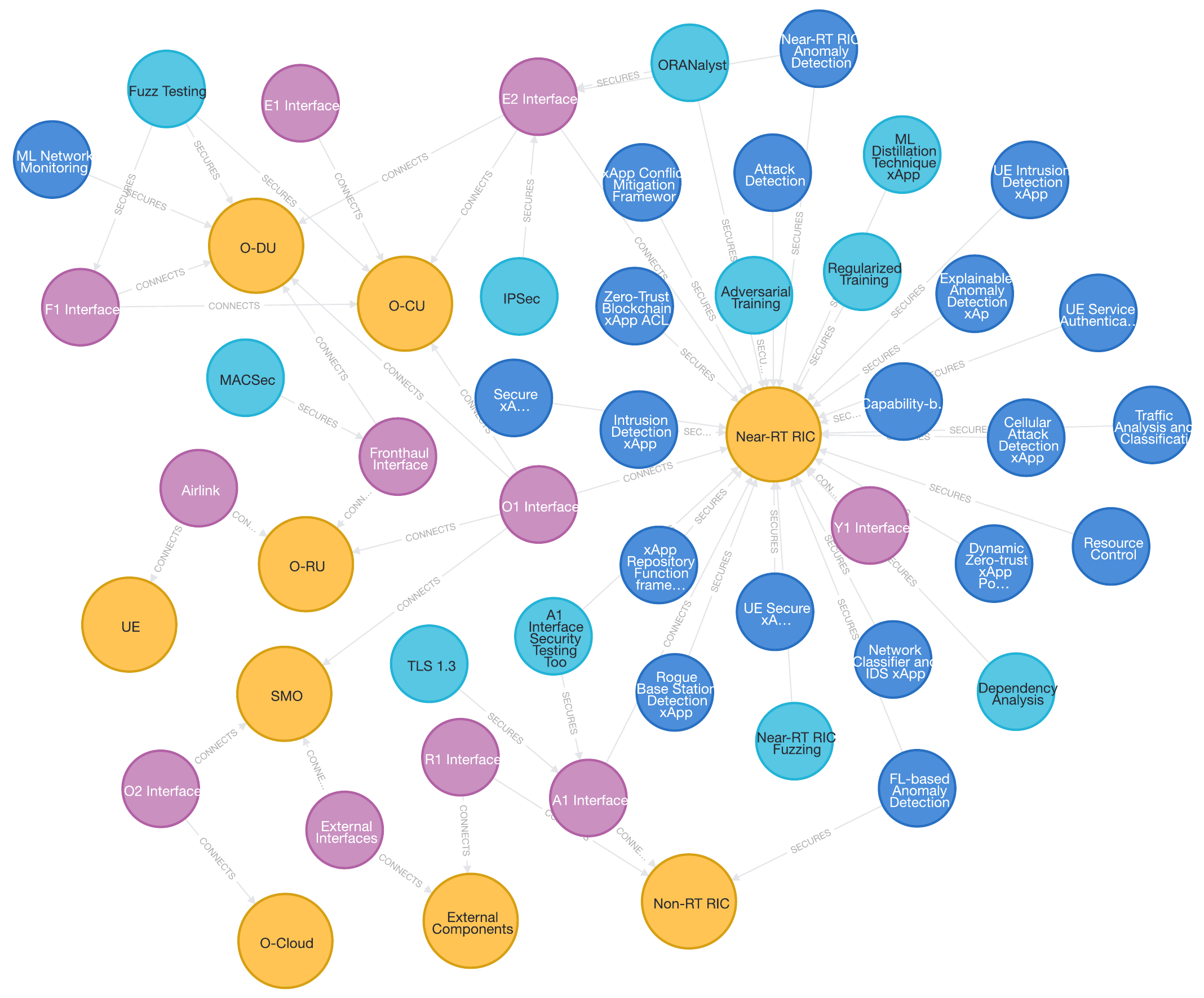}
  \caption{Map of Empirical Countermeasures for O-RAN: Prevention and detection-based countermeasures are shown as light blue (\lbdot) and dark blue (\dbdot), respectively.}
  \label{fig:mitigation-mapping}
\end{figure}

Figure~\ref{fig:experimental-attack} provides a graph-based representation of the O-RAN threats and their links to components and interfaces, resulting from querying our graph-based database for threats. We can directly extract that attacks are concentrated on the Near-RT RIC in terms of components and on the fronthaul and E2 in terms of interfaces.  Figure~\ref{fig:mitigation-mapping} provides the resulting map for prevention and detection mechanisms from a query on defenses. The vast majority of prevention and detection mechanisms is focused on the Near-RT RIC; interfaces are considered rather sporadically by mitigations. 

In Table~\ref{tab:threats}, we quantify the imbalance revealed by our graph analysis. While the Near-RT RIC and xApps are associated with over 70 distinct threats and numerous empirical studies, critical infrastructure components such as the SMO, Non-RT RIC, and O-Cloud have virtually no empirical security research despite their high theoretical risk scores. Furthermore, the E1 and F1 interfaces remain blind spots in the current literature. A primary driver of these gaps is experimental accessibility. Near-RT RIC and xApp studies benefit from widely used open-source reference implementations and publicly documented testbeds~\cite{osc, ric, oaiRAN}, which lowers the barrier to building reproducible experiments. In contrast, other components, like SMO and O-Cloud, require realistic and vendor-compatible deployments and are difficult to replicate with open-source or custom tools. This increases both cost and integration complexity relative to RIC- and interface-centric experiments. At the interface level, the limited coverage of E1/F1 may also reflect scoping conventions. E1 and F1 are primarily defined and deployed as part of the 3GPP CU/DU split and are not typically exposed through O-RAN-centric components, even though they remain security-relevant surfaces. Furthermore, the lack of UE-focused empirical work in the O-RAN literature likely reflects that UE security is often treated as orthogonal to O-RAN-specific components, as the UE has been explored extensively in 'conventional'  cellular architectures. However, experimental validation under O-RAN assumptions remains important (even for past vulnerabilities), because O-RAN introduces new control points that may influence UE and RAN behaviors. This uneven distribution highlights also a tendency in academia to focus on the novel ``AI-driven'' aspects of O-RAN, leaving the foundational management and orchestration layers underexplored and potentially exposed.

\begin{table}[!t]
 \footnotesize
  \caption{O-RAN Threats vs.~Empirical Studies: Comparison of threats from O-RAN specifications~\cite{oran2024threat} and existing work that introduce empirical security experiments (List~\ref{listing:empirical-work-table}).}
  \begin{tabularx}{\linewidth}{@{}l|l@{}c@{}X@{}}
  Type & Asset Name & \,Threats\, & \,References  \\
  \midrule
  \multirow[c]{10}{*}{\rotatebox{90}{Component}} & Near-RT RIC & 77 & %
  \!\!\cite{balakrishnanEnhancingORANSecurity2024,Sanaz2023BPM,Chiejina2024MLattack, Wen2024Spector, hungAnomalyDetectionMitigating2025, Thimmaraju2024a1interface, Yang2024osc, abdallaZTRANPrototypingZero2024, awadXAppsDDoSAttacks2024, eisoldtTrustworthyExecutionORAN2024, rumeshFederatedLearningAnomaly2024, adamczykConflictMitigationFramework2023, elhoudaBlockchainMeetsORAN2024, tsourdinisAIDrivenNetworkIntrusion2024, kouchakiAdvancingORANSecurity2024, wen6GXSecExplainableEdge2024, Atalay2023secure, groenTRACTORTrafficAnalysis2023, jiangOZTrustORANZeroTrust2023, Huang2023RBS,poisonXAppComNet2025,hybxlstmComputing2025,endlessSubscriptionsEuroSP2025,dynamicDefenseOJCOMS2025,ueMetricPoisoningCStdMag2025}
  \\
  \rotatebox{90}{} & Non-RT RIC & 75 & \!\!\cite{rumeshFederatedLearningAnomaly2024} \\
  \rotatebox{90}{} & O-Cloud & 72 & None in corpus \\
  \rotatebox{90}{} & External Compon. & 55 & None in corpus \\
  \rotatebox{90}{} & SMO & 48 & None in corpus \\
  \rotatebox{90}{} & O-DU & 47 &  \!\!\cite{groenTIMESAFETimingInterruption2024} \\
  \rotatebox{90}{} & O-RU & 43 & \!\!\cite{groenTIMESAFETimingInterruption2024} \\
  \rotatebox{90}{} & O-CU & 34 & \!\!\cite{dessources5GORANSecurity2024} \\
  \rotatebox{90}{} & UE & 26 & None in corpus \\

  \midrule

  \multirow[c]{11}{*}{\rotatebox{90}{Interface}} & Fronthaul Interface & 22 & \!\!\cite{Liao2022dos,Xing24OFH,groen2024interfaces, mufflerINFOCOM2025} \\
  \rotatebox{90}{} & A1 Interface & 8 & \!\!\cite{Thimmaraju2024a1interface,Tiberti2022mitma} \\
  \rotatebox{90}{} & R1 Interface & 7 & None in corpus \\
  \rotatebox{90}{} & O1 Interface & 3 & None in corpus \\
  \rotatebox{90}{} & E2 Interface & 2 & \!\!\cite{groen2024interfaces,Yang2024osc,hungAnomalyDetectionMitigating2025,kimSimulationARPSpoofing2024,Hung2024E2,layer2ProtectionORAN2025} \\
  \rotatebox{90}{} & O2 Interface & 2 & None in corpus \\
  \rotatebox{90}{} & Y1 Interface & 1 & None in corpus \\
  \rotatebox{90}{} & Airlink & 1 & None in corpus \\
  \rotatebox{90}{} & E1 Interface & 0 & None in corpus \\
  \rotatebox{90}{} & F1 Interface & 0 & \!\!\cite{dessources5GORANSecurity2024} \\
  \rotatebox{90}{} & External Interfaces & 0 & None in corpus \\
  \bottomrule
  \end{tabularx}
  \label{tab:threats}
  \end{table}

\subsection{Accuracy of LLM-based Automation}
\label{sec:llm-eval}

We evaluated the LLM outputs against a human-curated baseline. We excluded the \textit{Risks}/\textit{Threats} extraction as it was deterministically verified against the source schema, but focused on attacks, defenses, and preventive measures. Two authors independently labeled the LLM outputs, reconciling disagreements at the end of the analysis.

We classified the LLM outputs into two main groups:
\begin{enumerate}[leftmargin=*, itemsep=0pt, parsep=0pt, topsep=0pt]
\item \emph{Comparable:} The LLM output refers to the same phenomenon as the human baseline.
\item \emph{LLM-only:} The LLM identified a finding not present in the human baseline.
\end{enumerate}

For \emph{Comparable} items, we assessed their quality using an acceptability threshold. \emph{Exact} and \emph{Equivalent} matches are considered good. \emph{Sufficiently Equivalent} and \emph{Partially Equivalent} matches are considered acceptable---meaning they capture the core concept (e.g., correct attack type and target) but may omit minor auxiliary details or use slightly different terminology. All other labels (e.g., \emph{Insufficient/Vague}) are considered not acceptable.

Table~\ref{tab:results-summaries} summarizes our results. Panel (a) reports items that the LLM extracted \emph{and} that match a manual entry, grouped by equivalence to the manual reference. 
The thick rule marks the quality threshold (categories above it are considered suitable for our use). 
Panel (b) reports items surfaced \emph{only by the LLM} (no manual counterpart) and groups them by evidential support within the source paper. 
Counts are shown at right.

\textbf{Comparable Items} (Table~\ref{tab:results-summaries}a). 
The high proportion of \textit{Exact} and \textit{Equivalent} matches demonstrates strong semantic alignment between automated extraction and expert manual annotation. Even in cases classified as \textit{Sufficiently} or \textit{Partially Equivalent}, the LLM correctly identified the core security concept, with differences primarily attributable to terminology variation or abstraction level. Importantly, no comparable entries were classified below the \textit{Sufficiently Equivalent} threshold.

\begin{table}[t]
\centering
\caption{Comparison between the manual baseline and the LLM. The manual baseline contains \emph{attacks}, \emph{defenses}, and \emph{preventive measures} identified by human annotators. }
\captionsetup[subtable]{justification=raggedright,singlelinecheck=false}
\setlength{\tabcolsep}{3pt}

\begin{subtable}[t]{0.47\columnwidth}
\caption{Comparable with manual (54 in total)}
\label{tab:comp-manual}
\footnotesize
\renewcommand{\arraystretch}{1.06}
\begin{tabularx}{\linewidth}{@{}X r@{}}
\toprule
\textbf{Category} & \textbf{Count} \\
\midrule
\rowcolor{white}\textbf{TOTAL} & \textbf{51} \\
\rowcolor{gray!6} Exact & 37 \\
\rowcolor{gray!10} Equivalent & 11 \\
\rowcolor{gray!14} Sufficiently equivalent & 3 \\
\rowcolor{gray!18} Partially equivalent & 0 \\
\specialrule{.12em}{.25em}{.25em} %
\rowcolor{gray!30} Confounded (overlap w/ irrelevant) & 0 \\
\rowcolor{gray!50} Unsubstantiated / Hallucinated & 0 \\
\rowcolor{gray!60} Insufficient / Vague & 0 \\
\bottomrule
\end{tabularx}
\end{subtable}
\hfill
\begin{subtable}[t]{0.47\columnwidth}
\caption{LLM-only items (125 in total found by the LLM)}
\label{tab:not-comp-llm}
\footnotesize
\renewcommand{\arraystretch}{1.06}
\begin{tabularx}{\linewidth}{@{}X r@{}}
\toprule
\textbf{Category} & \textbf{Count} \\
\midrule
\rowcolor{white}\textbf{TOTAL} & \textbf{74} \\

\addlinespace[0.2em]
\multicolumn{2}{@{}l}{\textit{Direct LLM-only Matches}} \\
\rowcolor{gray!6} Supported (Explicit) & 16 \\
\rowcolor{gray!11} Supported (Implicit) & 13 \\

\specialrule{.12em}{.25em}{.25em}

\multicolumn{2}{@{}l}{\textit{Non-Direct LLM-only Matches}} \\
\rowcolor{gray!24} Over-fragmentation & 24 \\
\rowcolor{gray!24} Attack Used for Testing & 9 \\
\rowcolor{gray!24} Security Recommendation & 6 \\
\rowcolor{gray!24} Optimization/Deployment & 3 \\

\specialrule{.12em}{.25em}{.25em}

\multicolumn{2}{@{}l}{\textit{Extraction Errors}} \\
\rowcolor{gray!50} Unsupported (Hallucinated) & 2 \\
\rowcolor{gray!60} Not Related to O-RAN & 1 \\

\bottomrule
\end{tabularx}
\end{subtable}

\vspace{0.35em}
\begin{minipage}{\columnwidth}
\centering
\begin{tikzpicture}[x=\columnwidth,y=1cm]
  \def\W{0.94}
  \def\H{0.18}
  \shade[left color=gray!10,right color=gray!70] (0,0) rectangle (\W,\H);
  \draw[black!30,line width=0.2pt] (0,0) rectangle (\W,\H);
  \draw[black!35,line width=0.2pt] (0,0) -- (0,\H);
  \draw[black!35,line width=0.2pt] (0.5*\W,0) -- (0.5*\W,\H);
  \draw[black!35,line width=0.2pt] (\W,0) -- (\W,\H);
  \node[anchor=west] at (0,-0.14) {\scriptsize High alignment};
  \node              at (0.5*\W,-0.14) {\scriptsize Moderate};
  \node[anchor=east] at (\W,-0.14) {\scriptsize Low alignment};
\end{tikzpicture}
\end{minipage}
\label{tab:results-summaries}
\end{table}

\textbf{LLM-only Items} (Table~\ref{tab:results-summaries}b). 
These items were analyzed further and grouped based on evidential support and structural characteristics. Within this group, we distinguish between:
\begin{itemize}[leftmargin=*, itemsep=0pt, parsep=0pt, topsep=0pt]
\item \textbf{Direct Matches:} Findings that are clearly supported by the source paper and represent legitimate standalone mechanisms not captured during manual annotation.
\item \textbf{Non-Direct Matches:} Findings that are generally supported in the source paper but differ in granularity, scope, or abstraction from the manual baseline.
\end{itemize}

\noindent Among the \textit{Non-Direct Matches} (42 entries), we distinguish:
\begin{itemize}[leftmargin=*, itemsep=0pt, parsep=0pt, topsep=0pt]
\item \textbf{Over-fragmentation (24 entries):} These cases occur when the LLM extracts internal sub-components, implementation steps, or supporting mechanisms of a broader attack or defense as standalone entries. While conceptually correct and supported in the paper, they belong to a higher-level mechanism already captured in the manual baseline.
\item \textbf{Attack Used for Testing (9):} Refers to known attack techniques that were employed to evaluate the robustness of proposed defenses. They are valid and supported in the paper, but are not novel attack contributions on their own.
\item \textbf{Security Recommendation (6)}: Refers to defenses or mitigation strategies mentioned in discussion/conclusion sections but not implemented or experimentally validated.
\item \textbf{Optimization/Deployment (3):} Corresponds to configuration choices, parameter tuning, deployment refinements, or engineering decisions that were implemented and evaluated in the paper. While technically correct and supported, they do not introduce new security guarantees or conceptual mechanisms, but  describe how a defense, attack, or preventive measure was deployed or optimized.
\end{itemize}

\textbf{Takeaways.}
The results show that LLM-based automation is effective for scaling security knowledge extraction. The high proportion of \textit{Exact} and \textit{Equivalent} matches demonstrates strong semantic alignment with expert annotations, while LLM-only findings improve coverage and recall.
Most non-direct LLM-only items (e.g., over-fragmentation, testing artifacts, deployment optimizations) do not introduce conceptual errors. Instead, they reflect finer-grained decomposition or contextual details already implicit at higher abstraction levels. In practice, the LLM expands the representation rather than distorting it.
However, expert oversight remains essential. We position the LLM as a high-recall assistant within a human-in-the-loop workflow. As described in Section~\ref{subsec:update-database}, our GUI enables users to inspect, validate, merge, or discard extracted entries before integrating them into the graph.

%% file: stakeholder_insights.tex
\section{Insights for O-RAN Stakeholders}
\label{sec:insights}

In this section, we demonstrate the practical utility of our graph-based framework. By querying the database, we derive stakeholder-specific insights that would be difficult to synthesize from static documents alone.
The Appendix includes references to all queries leveraged.

\subsection{Insights for Academia}
\label{subsec:insights-academia}

\noindent\textbf{(Question)}
\noindent\textit{Where are the critical gaps in empirical O-RAN security research?}

\noindent\textbf{(Method)}
We break this into two criteria: \textbf{1)} Understanding the current O-RAN security research landscape, and \textbf{2)} revealing what components are most in need of scrutiny. 
We dissect the former by constructing a query that enumerates attacks, defensive measures, and preventive measures as total quantities, as well as distinct works (List~\ref{listing:insight-academic-4}). 
We investigate the latter with queries that search for components and interfaces with either no attacks, no defenses, or no preventative measures (List~\ref{listing:insight-academia-2}). To elicit additional insights, we integrate threat information to prioritize which components should be investigated first, matching threats to our components. We then further refine our query, matching only direct and high-risk threats to these components. Lastly, we introduce a query that returns components or interfaces without documented threats to elicit potential gaps in the O-RAN threat model~\cite{oran2024threat} (List~\ref{listing:insight-academia-8}).

\noindent\textbf{(Insight)}
Our analysis reveals a stark imbalance. Research is heavily skewed toward the "Intelligent Control" layer: the Near-RT RIC and xApps are well-covered. 
In contrast, critical infrastructure components are neglected.
For example, we find a total of 32 attacks, 32 defenses, and 12 preventive measures. However, as many works propose multiple attacks or defenses, we deduplicate these, revealing only 18 attack papers, 21 defense papers, and 9 preventive-measure papers. The $43$\% decrease from attacks to distinct attack papers indicates that individual papers tend to propose multiple attacks, whereas defense and preventive-measure papers do so less frequently.
Lastly, as our literature review suggested a concentration on xApps and Near-RT RIC, we further refine our query to exclude these components. This final subset yields 7 attacks, 5 defense, and 3 preventive measures. Moreover, 3 of the 7 attacks and the 2 defense concentrate their focus on the Fronthaul Interface~\cite{groenTIMESAFETimingInterruption2024,Liao2022dos,groen2024interfaces,mufflerINFOCOM2025}, leaving other components with little to no empirical security research.
For attacks, the O-Cloud (55), Non-RT RIC (47), SMO (39), External Components (39), and O-CU (27) are the components that require the most empirical attack work. Rigorous attacks are crucial to providing defenses with real-world scenarios to evaluate against. 
Defenses and preventative measures overlap with attack component priorities for the first four components, but also require defense work in the O-RU (25) and UE (24).
Our final query sheds light upon two interfaces that do not have any documented potential threats in the O-RAN threat model~\cite{oran2024threat}: the E1 Interface and the F1 Interface. Work that investigates the security of these interfaces would likely be highly valuable to the O-RAN community and the security working group. Threats demonstrated here should be submitted to the O-RAN Alliance for inclusion in the threat model, as prior work has done~\cite{groenTIMESAFETimingInterruption2024,Thimmaraju2024a1interface}. 

\noindent\textbf{(Recommendation)} Future academic work should pivot away from redundant ML adversarial examples and focus on the "plumbing" of O-RAN, specifically the O-DU and internal interfaces where theoretical risks remain unvalidated.

\begin{table*}[!t]
\caption{CWE breakdown by vulnerability class and impact for O-RAN software. %
}
\centering
\small
\setlength{\tabcolsep}{8pt}
\renewcommand{\arraystretch}{0.3}
\begin{tabular}{@{}p{1.95cm} p{8.25cm} p{6.4cm}@{}}
\toprule
\textbf{Class} & \textbf{CWEs (name, count)} & \textbf{Impact} \\
\midrule
Memory Safety &
CWE-787 Out-of-bounds Write (3), CWE-129 Improper Array Index Validation (3),
CWE-120 Buffer Overflow (2), CWE-125 Out-of-bounds Read (2),
CWE-476 NULL Dereference (1) &
Crash/DoS. Memory corruption. Potential Remote Code Execution (context-dependent). \\
\midrule
Resource Exhaustion &
CWE-400 Uncontrolled Resource Consumption (3),
CWE-770 Unbounded Resource Allocation (1),
CWE-835 Infinite Loop (1) &
Slow DoS via CPU/memory starvation and tail-latency inflation (harms near-RT control). \\
\midrule
Assertions &
CWE-617 Reachable Assertion (3) &
Crashable DoS via unexpected input/state reaching production asserts. \\
\midrule
Access Control  &
CWE-862 Missing Authorization (1) &
Unauthorized privileged actions. \\
\midrule
Input Semantics  &
CWE-20 Improper Input Validation (1) &
Invalid-but-parsable inputs cause incorrect control actions that may cascade into DoS/memory faults. \\
\bottomrule
\end{tabular}
\label{tab:oran-cwe-impact}
\end{table*}

\subsection{Insights for Operators}

\noindent\textbf{(Question)} \textit{What strategies have proven most effective to uncover vulnerabilities in O-RAN software?}

\noindent\textbf{(Method)}
To answer this, we first query for CVEs attributed to specific works and the number of CVEs each work disclosed (List~\ref{listing:insight-academia-5}). We also query for CVE attribution to identify the individuals who were lead authors or security researchers who disclosed the vulnerabilities (List~\ref{listing:insight-academia-6}).

\noindent\textbf{(Insights)}
Our results reveal that out of all 27 CVEs, 23 were discovered by only four works~\cite{hungAnomalyDetectionMitigating2025, Hung2024E2, Yang2024osc,endlessSubscriptionsEuroSP2025}, while the remaining four are currently unassociated with a paper. Investigating by lead author or discloser rather than papers, we find that three security researchers are responsible for 24 CVEs. This high concentration of individuals and papers alludes to the nascency of the O-RAN software ecosystem, as well as the need to expand the number of security researchers seeking O-RAN vulnerabilities, potentially through crowd-sourced programs, such as bug bounties~\cite{maillartGivenEnoughEyeballs2017}.
Interestingly, two of the four relevant works use fuzzing as their technique of choice~\cite{hungAnomalyDetectionMitigating2025,Yang2024osc}. These two papers account for 18 of the 20 paper-associated CVEs (90\%).
Fuzzing has proven itself a rigorous technique to discover bugs~\cite{schloegelSoKPrudentEvaluation2024}. 
Software with no disclosed CVEs currently includes the O-RAN Software Community AIMLFW, NONRTRIC, INF, OAM, SMO, ODUHIGH, ODULOW, ORU, OCU, along with the OpenAirInterface oai-cn5g, oai-ran, and lastly the srsRAN gNB project (List~\ref{listing:insight-academia-7}).

\subsection{Insights for Governments and Regulators}
\label{subsec:insights-governments}

\noindent\textbf{(Question)}
\noindent\textit{What components pose the greatest risk and might become a nation-state target to disrupt cellular infrastructure amid escalating geopolitical tensions?} %

\noindent\textbf{(Method)} To understand nation-state availability attacks, we design a query that matches threats that explicitly mention components and interfaces. We then further refine our search to availability-specific threats, which were ranked as high-severity but with low likelihood. We consider these characteristics to be consistent with a threat actor who can invest significant resources to execute a highly complex attack with significant ramifications. 
We finally appended two optional matches to return attacks from research, as well as relevant software that requires scrutiny, for added context (List~\ref{listing:insight-government-1}).

\noindent\textbf{(Insights)} The top components and their threat degrees for availability attacks are the O-DU (4), O-RU (4), Non-RT RIC (2), and Near-RT RIC (2). O-DU and O-RU threats include unauthorized access to the layer 1 OFH interfaces, and rogue Precision Time Protocol (PTP) spoofing, intercepting, or becoming the grand master clock. The query result also includes the relevant work that introduced these threats into the threat model~\cite{groenTIMESAFETimingInterruption2024}. The Near-RT and Non-RT RICs share A1-interface access threats and can also suffer denial-of-service through indirect abuse of A1 policy and feedback channels.

\noindent\textbf{(Recommendation)} Governments and Regulators should prioritize physical security and tamper-proofing requirements for O-DU/O-RU deployments, as these are the nexus for catastrophic availability attacks.

\subsection{Insights for O-RAN Software Developers}

\noindent\textbf{(Question)}
\noindent\textit{What root causes drive the majority of O-RAN vulnerabilities?}

\noindent\textbf{(Method)}
To answer this, we build a query that matches all CWEs with associated CVEs from O-RAN software and returns the degree count (List~\ref{listing:insight-operator-2}). 

\noindent\textbf{(Insights)}
After retrieving the CWEs, we grouped them into high-level vulnerability classes and summarize the results in Table~\ref{tab:oran-cwe-impact}. Memory-safety issues dominate the observed anti-patterns, with frequent instances of improper array index validation (3), out-of-bounds writes (3), and classic buffer overflows (2). This concentration suggests that a large fraction of O-RAN risk currently sits in input-facing parsing and state-handling code, where malformed or adversarial inputs can trigger memory violations. Beyond memory safety, resource-exhaustion issues are also common, indicating that availability threats are practical in O-RAN software with some vulnerabilities enabling ``fast DoS'' via crashes, while others enabling ``slow DoS'' through CPU/memory starvation and tail-latency inflation. Finally, the presence of reachable assertions (CWE-617) highlights crashable code paths that are still exposed to unexpected input or state transitions, pointing to the need for stronger negative testing of protocol handlers and state machines.
If this trend continues as O-RAN software matures, memory vulnerabilities are likely to plague O-RAN software, as they have comprised 70\% of security vulnerabilities in operating systems and browsers~\cite{kimAreWeDone2021}. Using memory-safe languages, such as Rust or Python, where appropriate, can help alleviate these vulnerabilities, making O-RAN software more secure, in line with industry recommendations~\cite {cisaMem2,cisaMem3}. The ORAN-SC-RIC~\cite{ric} has made progress in this direction, introducing a Python API with C bindings to enable developers to safely interface with the RIC from an xApp. Developers should also consider introducing a roadmap to memory safety, to better mitigate these concerns over the long term~\cite{cisaMem1}.

\noindent\textbf{(Recommendation)} The O-RAN ecosystem faces a systemic memory safety risk. Developers must prioritize the adoption of memory-safe languages (like Rust or Go) for new components. For existing C++ codebases (like the RIC), enabling modern mitigations (ASLR, DEP, Stack Canaries) is a non-negotiable baseline requirement. Additionally, the software must be subject to continuous security testing (e.g., negative testing and fuzzing) in order to identify crashable paths early and prevent malformed or adversarial inputs from reaching unsafe parsing and state-machine logic. 
Finally, enforcement of proper access control is needed for privileged operations (e.g., authZ-by-default with least privilege), alongside thorough semantic validation (ranges, cross-field consistency, and valid state transitions) before applying control actions.

%% file: future_research_and_conclusion.tex
\section{Future Research Directions}
Our ontological security database is designed to evolve beyond O-RAN and support broader specifications and domains. The graph-based analysis (Sec.~\ref{sec:insights}) highlights structural gaps in the current O-RAN security landscape. In particular, defensive mechanisms significantly outnumber demonstrated attacks, indicating the need for deeper exploration of novel attack vectors. Research efforts are also unevenly distributed: components such as the Near-RT RIC receive substantial attention, while critical infrastructure elements (e.g., O-Cloud, Non-RT RIC, SMO) remain comparatively understudied despite their potential impact. Notably, interfaces such as E1 and F1 currently lack documented threats in the O-RAN threat model~\cite{oran2024threat}.

\noindent\textbf{Extensibility:}
Given the rapid evolution of O-RAN specifications and research, the database is designed for straightforward extension. Its spreadsheet-backed ground truth and GUI-supported workflow (Sec.~\ref{subsec:update-database}) allow users to incorporate new attacks, defenses, and components with minimal friction. We advocate maintaining the dataset in a public repository to facilitate community-driven updates as new findings emerge.

\section{Conclusion}

In this paper, we introduced a graph-based security tool for O-RAN that enables an ontological perspective. Our tool distills information across sources from industry, academia, and government into a common analytical representation. This tool enables a guided approach to identifying important risks in O-RAN networks and directions for future work. Complementing expert curation, our methodology integrates a schema-constrained LLM extraction module that generates machine-checkable, ontology-aligned records; together with an easy-to-use GUI, this supports reproducible, human-supervised updates as specifications and the literature evolve. We believe our security investigation and contributions will help enhance O-RAN security and encourage other researchers to pursue further studies in this area.

%% file: Appendix.tex
\section*{Appendix}

\subsection{Access to Graph Database}
\label{appendix:database-access}
\begin{figure}[b]
\centering
    \begin{subfigure}{.3\textwidth}
        \centering
        \includegraphics[width=\textwidth]{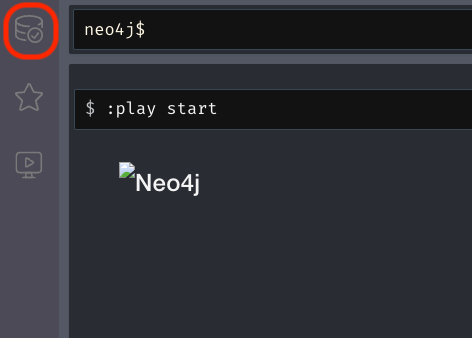}
        \caption{Enumerate the database schema.}
    \end{subfigure}
    \begin{subfigure}{.3\textwidth}
        \centering
        \includegraphics[width=\textwidth]{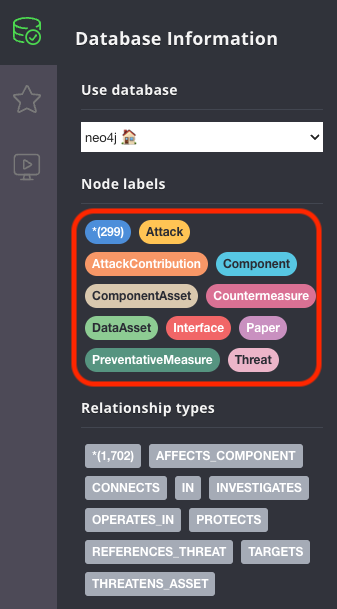}
        \caption{Query for various components.}
    \end{subfigure}
    \begin{subfigure}{.3\textwidth}
        \centering
        \includegraphics[width=\textwidth]{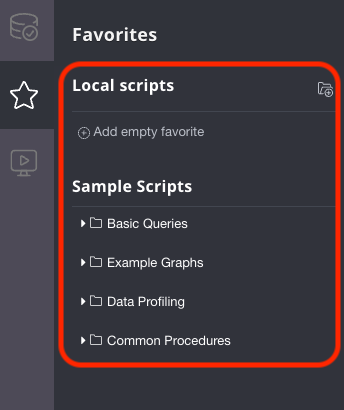}
        \caption{Execute saved queries.}
    \end{subfigure}
    \caption{Graph Database Navigation. Double-clicking on a returned component in the view retrieves connecting nodes.}
    \label{appendix:graph-database-screenshots}
\end{figure}

The artifact repository contain deployment instructions and document supported browsers. See Fig.~\ref{appendix:graph-database-screenshots} for guidance on navigating the database browser.

\subsection{Discovered Typographical Errors}
\label{appendix:typos}

Table \ref{table:typos} lists the typographical errors we identified.

\begin{table}[htb]
\caption{Typographical errors discovered in the O-RAN specifications~\cite{ORANSecurityThreatModel} (\texttt{\small{R003-v03.0}}) after database ingestion.}
\small
\centering
\resizebox{1.\columnwidth}{!}{%
\begin{tabularx}{\linewidth}{@{}l@{}l@{}l@{}}
\toprule
\textbf{Page} & \textbf{Description} & \textbf{Example} \\
\midrule
90 & Missing dash in asset ID & \scriptsize{\texttt{ASSET D-11 $\rightarrow$ ASSET-D-11}} \\
92 & ``Framework'' misspelled & \scriptsize{\texttt{Fr\st{o}amework $\rightarrow$ Framework}} \\
92, 93, 96\,\, & Underscore instead of dash\,\, & \scriptsize{\texttt{ASSET-C\_21 $\rightarrow$ ASSET-C-21}} \\
24, 102 & Inconsistent capitalization & \scriptsize{\texttt{xAPP $\rightarrow$ xApp}} \\
24, 26 & Inconsistent capitalization & \scriptsize{\texttt{rAPP $\rightarrow$ rApp}} \\
90, 91 & Undefined asset IDs & \scriptsize{\texttt{ASSET-D-41, ASSET-D-43}} \\
\bottomrule
\end{tabularx}
}
\label{table:typos}
\end{table}

\subsection{Example Queries}

Listings~\ref{listing:papers-count}-\ref{listing:insight-academic-4} provide example queries.

\begin{figure}[tb]
    \begin{minipage}{\linewidth}
    \begin{lstlisting}[style=Text,breaklines, frame=single, caption={Queries used to quantify O-RAN security literature through Scopus.},captionpos=b, label=listing:papers-count]
\bScopus Query: \ubTITLE-ABS-KEY ("ORAN" OR "O-RAN" OR "Open RAN" OR "Open Radio Access Network") AND TITLE-ABS-KEY ("Security")
\end{lstlisting}
    \end{minipage}
\end{figure}

\begin{figure}[tb]
    \begin{minipage}{\linewidth}
    \begin{lstlisting}[style=CStyle,breaklines, frame=single, caption={Returns components and interfaces with either no attacks, defenses, or preventive measures. Rows are sorted in descending order of a high-risk threat degree.},captionpos=b, label=listing:insight-academia-2]
MATCH (n)
WHERE (n:Component OR n:Interface)
    // Toggle filters below as necessary
    AND NOT ((n)<-[:TARGETS]-(:Attack))
//   AND NOT ((n)<-[:SECURES]-(:Defense))
//   AND NOT ((n)<-[:SECURES]-(:PreventiveMeasure))
OPTIONAL MATCH (t:Threat)-[:TARGETS]->(n)
WHERE t.risk_score = 'High'
RETURN n.name AS component, count(DISTINCT t) AS threatDegree
ORDER BY threatDegree DESC
\end{lstlisting}
    \end{minipage}
\end{figure}

\begin{figure}[tb]
    \begin{minipage}{\linewidth}
    \begin{lstlisting}[style=CStyle,breaklines, frame=single, caption={Returns components and interfaces with no documented threats.},captionpos=b, label=listing:insight-academia-8]
MATCH (c:Component)
WHERE NOT (c)<-[:TARGETS]-(:Threat)
RETURN c.name AS name
UNION
MATCH (i:Interface)
WHERE NOT (i)<-[:TARGETS]-(:Threat)
RETURN i.name AS name      
\end{lstlisting}
    \end{minipage}
\end{figure}

\begin{figure}[tb]
    \begin{minipage}{\linewidth}
    \begin{lstlisting}[style=CStyle,breaklines, frame=single, caption={Returns a list of unique works mapped to the number of CVEs each discovered.},captionpos=b, label=listing:insight-academia-5]
MATCH (c:CVE)
UNWIND c.reference AS ref
WITH toString(ref) AS refString
WHERE refString <> 'NaN'
RETURN refString AS `CVE Reference`, count(*) AS `Number of CVEs Discovered`
\end{lstlisting}
    \end{minipage}
\end{figure}

\begin{figure}[tb]
    \begin{minipage}{\linewidth}
    \begin{lstlisting}[style=CStyle,breaklines, frame=single, caption={Returns a list of attributed persons mapped to the number of CVEs each discovered.},captionpos=b, label=listing:insight-academia-6]
MATCH (c:CVE)
UNWIND c.attribution AS ref
WITH toString(ref) AS refString
WHERE refString <> 'NaN'
RETURN refString AS `Attribution`, count(*) AS `Number of CVEs Discovered`
\end{lstlisting}
    \end{minipage}
\end{figure}

\begin{figure}[tb]
    \begin{minipage}{\linewidth}
    \begin{lstlisting}[style=CStyle,breaklines, frame=single, caption={Returns a list of software with no discovered CVEs. Rows are sorted according to number of threats attacking the component each implements.},captionpos=b, label=listing:insight-academia-7]
MATCH (s:Software)
WHERE NOT ((:CVE)-[:AFFECTS]->(s))
MATCH (s)-[:IMPLEMENTS]->(c:Component)
OPTIONAL MATCH (t:Threat)-[:TARGETS]->(c)
WITH s, c, COUNT(t) AS threatCount
RETURN s.name AS software, c.name AS component, threatCount
ORDER BY threatCount DESC;        
\end{lstlisting}
    \end{minipage}
\end{figure}

\begin{figure}[tb]
    \begin{minipage}{\linewidth}
    \begin{lstlisting}[style=CStyle,breaklines, frame=single, caption={Returns targets that specifically target components or interfaces. Threats must be rated as high severity but low likelihood. Map attacks from literature and implementing software. Sort in descending order by threat degree.},captionpos=b, label=listing:insight-government-1]
MATCH (t:Threat)-[:TARGETS { all: false }]->(n)
WHERE t.severity_level = 'High'
  AND t.likelihood_level = 'Low'
  AND t.availability = true
  AND (n:Component OR n:Interface)
OPTIONAL MATCH (a:Attack)-[:TARGETS]->(n)
OPTIONAL MATCH (s:Software)-[:IMPLEMENTS]->(n)
RETURN n.name AS TargetEntity,
       count(DISTINCT t) AS threatDegree,
       collect(DISTINCT t.threat_title) AS threatTitles,
       collect(DISTINCT a.name) AS attackNames,
       collect(DISTINCT s.name) AS softwareNames
ORDER BY threatDegree DESC
\end{lstlisting}
    \end{minipage}
\end{figure}

\begin{figure}[tb]
    \begin{minipage}{\linewidth}
    \begin{lstlisting}[style=CStyle,breaklines, frame=single, caption={Returns all CWEs and CVEs, sorted in descending order by the number of CVEs associated with each CWE.},captionpos=b, label=listing:insight-operator-2]
MATCH (c:CWE)<-[:ASSOCIATED_WITH]-(cve:CVE)
RETURN c.name AS `CWE`, c.description AS `Description`, count(cve) AS `Degree Count`
ORDER BY `Degree Count` DESC
\end{lstlisting}
    \end{minipage}
\end{figure}

\begin{figure}[tb]
    \begin{minipage}{\linewidth}
    \begin{lstlisting}[style=CStyle,breaklines, frame=single, caption={Returns components and interfaces, matched with associated threats, attacks, defenses, and preventive measures. Sort results in descending order by threat degree.},captionpos=b, label=listing:empirical-work-table]
MATCH (n)
WHERE n:Component OR n:Interface
OPTIONAL MATCH (n)<-[:TARGETS]-(t:Threat)
WITH n, count(DISTINCT t) AS threat_degree
OPTIONAL MATCH (n)<-[rel:TARGETS|SECURES]-(ent)
WHERE ent:Attack OR ent:Defense OR ent:PreventiveMeasure
WITH n, threat_degree, collect(DISTINCT ent.reference) AS refs
RETURN n.name AS name,
       threat_degree,
       [r IN refs WHERE r IS NOT NULL] AS refs
ORDER BY threat_degree DESC; 
\end{lstlisting}
    \end{minipage}
\end{figure}

\begin{figure}[tb]
    \begin{minipage}{\linewidth}
    \begin{lstlisting}[style=CStyle,breaklines, frame=single, caption={Returns the number of distinct attacks, defenses, and preventive measures as works, total proposed, and proposed that are not related to the Near-RT RIC. },captionpos=b, label=listing:insight-academic-4]
MATCH (n:Attack)
WITH "Attack Works" AS type, count(DISTINCT n.reference) AS count,
        collect(DISTINCT n.reference) AS references
RETURN type, count, references
UNION
MATCH (n:Defense)
WITH "Defense Works" AS type, count(DISTINCT n.reference) AS count,
        collect(DISTINCT n.reference) AS references
RETURN type, count, references
UNION
MATCH (n:PreventiveMeasure)
WITH "PreventiveMeasure Works" AS type, count(DISTINCT n.reference) AS count,
        collect(DISTINCT n.reference) AS references
RETURN type, count, references
UNION
MATCH (n:Attack)
WITH "Attacks Total" AS type, count(n.reference) AS count,
        collect(n.reference) AS references
RETURN type, count, references
UNION
MATCH (n:Defense)
WITH "Defenses Total" AS type, count(n.reference) AS count,
        collect(n.reference) AS references
RETURN type, count, references
UNION
MATCH (n:PreventiveMeasure)
WITH "PreventiveMeasures Total" AS type, count(n.reference) AS count,
        collect(n.reference) AS references
RETURN type, count, references
UNION
MATCH (n:Attack)
WHERE NOT ((n)-[:TARGETS]->(:Component {name: "Near-RT RIC"}))
  AND NOT ((n)-[:TARGETS]->(:Interface {name: "E2 Interface"}))
WITH "Attacks (Not Near-RT RIC)" AS type, count(DISTINCT n.reference) AS count,
     collect(DISTINCT n.reference) AS references
RETURN type, count, references
UNION
MATCH (n:Defense)
WHERE NOT ((n)-[:SECURES]->(:Component {name: "Near-RT RIC"}))
  AND NOT ((n)-[:SECURES]->(:Interface {name: "E2 Interface"}))
WITH "Defenses (Not Near-RT RIC)" AS type, count(DISTINCT n.reference) AS count,
     collect(DISTINCT n.reference) AS references
RETURN type, count, references
UNION
MATCH (n:PreventiveMeasure)
WHERE NOT ((n)-[:SECURES]->(:Component {name: "Near-RT RIC"}))
  AND NOT ((n)-[:SECURES]->(:Interface {name: "E2 Interface"}))
WITH "PreventiveMeasure (Not Near-RT RIC)" AS type, count(DISTINCT n.reference) AS count,
     collect(DISTINCT n.reference) AS references
RETURN type, count, references
\end{lstlisting}
    \end{minipage}
\end{figure}

\clearpage